\documentclass[aps,prd,reprint,showpacs,superscriptaddress,amsmath,amssymb,amsfonts,floatfix]{revtex4-1}  
\pdfoutput=1
\usepackage{graphicx}
\usepackage{bm}

\newcommand{\beq}{\begin{equation}}
\newcommand{\beqa}{\begin{eqnarray}}
\newcommand{\bey}{\begin{eqnarray}}
\newcommand{\eeq}{\end{equation}}
\newcommand{\eey}{\end{eqnarray}}
\newcommand{\eeqa}{\end{eqnarray}}

\def\lsim{\mathrel{\raise.3ex\hbox{$<$\kern-.75em\lower1ex\hbox{$\sim$}}}}
\def\gsim{\mathrel{\raise.3ex\hbox{$  $\kern-.75em\lower1ex\hbox{$\sim$}}}}

\begin{document}

\title{Testing quasilinear modified Newtonian dynamics in the Solar System}
\author{Pasquale Galianni}
\email[Electronic address: ]{pg25@st-andrews.ac.uk}
\affiliation{SUPA, School of Physics \& Astronomy, University of St Andrews, KY16 9SS, United Kingdom}
\author{Martin Feix}
\affiliation{Department of Physics, Technion - Israel Institute of Technology, Technion City, 32000 Haifa, Israel}
\author{Hongsheng Zhao}
\affiliation{SUPA, School of Physics \& Astronomy, University of St Andrews, KY16 9SS, United Kingdom}
\author{Keith Horne}
\affiliation{SUPA, School of Physics \& Astronomy, University of St Andrews, KY16 9SS, United Kingdom}

\begin{abstract}
A unique signature of the modified Newtonian dynamics (MOND) paradigm is its peculiar behavior in the vicinity of the points where 
the total Newtonian acceleration exactly cancels. In the Solar System, these are the saddle points of the gravitational potential
near the planets. Typically, such points are embedded into low-acceleration bubbles where modified gravity theories \`a la MOND predict
significant deviations from Newton's laws. As has been pointed out recently, the Earth-Sun bubble may be visited by the
LISA Pathfinder spacecraft in the near future, providing a unique occasion to put these theories to a direct test. In this work, we present
a high-precision model of the Solar System's gravitational potential to determine accurate positions and motions of these saddle points and
study the predicted dynamical anomalies within the framework of quasi-linear MOND. Considering the expected sensitivity of the LISA Pathfinder
probe, we argue that interpolation functions which exhibit a ``faster'' transition between the two dynamical regimes have a good chance of
surviving a null result. An example of such a function is the QMOND analog of the so-called simple interpolating function which agrees well
with much of the extragalactic phenomenology. We have also discovered that several of Saturn's outermost satellites periodically intersect the
Saturn-Sun bubble, providing the first example of Solar System objects that regularly undergo the MOND regime.
\end{abstract}
\maketitle

\section{Introduction}
\label{section1}
Since the beginning of the past century, the Solar System (SS) has been a major lab for testing the validity of general relativity (GR). Ranging from the first evidence of Mercury's perihelion precession \cite{leverrier} up to the recent tests of the Lense-Thirring effect \cite{ciufolini,everitt}, the SS has played a key role in our understanding of the laws of gravity. All SS tests of GR performed so far, however, have tested Einstein's theory in the regime of moderately strong gravity, i.e. where the total Newtonian acceleration exceeds the typical values found at galactic and cosmological scales by several orders of magnitude.
At larger scales, however, it is well-known that GR fails to account for the dynamics of galaxies and galaxy clusters unless a massive component of dark matter (DM) \cite{zwicky,spergel2003} is taken into account. The current lack of knowledge about the composition and behavior of this dark component of the Universe, which has recently increased with the study of the internal dynamics of dwarf galaxies \cite{penarrubia}, together with the failing attempts to detect DM particles \cite{Bertone2005}, motivates research on different alternative gravity theories such as $f(R)$ gravity \cite{fr}, conformal gravity \cite{mannheim} and various formulations of the modified Newtonian dynamics (MOND) paradigm \cite{Mond3,Mond2,Mond1,zhao2007coin} like, for instance, tensor-vector-scalar theory (TeVeS)
\cite{bekenstein2004} and quasi-linear MOND \cite{qmond}.

Alternative frameworks of the latter kind predict important deviations from Newtonian mechanics in regions where the total gravitational acceleration is much smaller than the value of $a_{0}\approx 10^{-10}$ m s$^{-2}$, i.e. $g_{N} \ll a_{0}$. In most parts of the SS, the total gravitational acceleration exceeds this value by several orders of magnitude. As first noted in Ref. \cite{bekenstein2006}, however, there exist numerous points in the SS where the sum of all gravitational fields exactly cancels out. These are the saddle points (SPs) of the Newtonian potential near the planets which are close to, but do not coincide with the Lagrangian L1 points (where also the centrifugal potential is taken into account). These SPs are embedded into ellipsoidal low-acceleration regions (bubbles) where the gradient of the gravitational potential becomes very small, i.e. comparable to or smaller than $a_{0}$, providing an ideal environment for testing some of the predictions of this set of modified gravity theories.

The LISA Pathfinder space mission (LPF) \cite{lisa} is an ESA project currently planned for launch in 2014. LPF will test technologies that will be employed in the following LISA (Laser Interferometer Space Antenna) mission, which aims at the detection of gravitational waves. The LPF will consist of two closely spaced test masses, whose distance relative to each other will be measured by means of an interferometer. The spacecraft will be shielded from any non-gravitational influence such that the masses inside the spacecraft will be in an almost perfect geodesic motion. The LPF will therefore be able to measure tidal stresses to an unprecedented accuracy. The expected trajectory of LPF will be a Lissajous orbit around the L1 point. With minimum corrections to its planned orbit, LPF may be redirected to the low-gravity region encasing the Earth-Sun or Moon-Sun SPs \cite{bevis2010}. In previous works, it has been claimed that the LPF instruments have enough sensitivity to detect the extra tidal stress signal predicted by MOND and TeVeS \cite{bekenstein2006, bevis2010}. As such, this particular mission is believed to provide a unique occasion to put such modified gravity frameworks to a direct test. The strength of the MONDian perturbations that the spacecraft will experience when crossing the Earth-Sun low-gravity bubble, however, depends strongly on the transition speed between the Newtonian and MOND regimes, as well as on the intrinsic structure of whichever MOND formulation is used. The full space of possible results has not yet been explored in detail. 
In the present contribution, we use the recently formulated quasi-linear MOND framework (QMOND) \cite{qmond} to make predictions about the expected deviations from Newtonian gravity that would occur inside the low-gravity region encasing the Earth-Sun and Moon-Sun SPs. 
Since the positions and sizes of the low-acceleration bubbles depend on the specific configuration of the SS at a certain instant, we also address the problem of determining their positions to very high precision. This is a necessary prerequisite to any realistic performance of a saddle-flyby experiment.

The present paper is structured as follows:
In Sec. \ref{section2}, we introduce the basic concepts regarding QMOND and low-acceleration bubbles. In Sec. \ref{section3}, we present an accurate model of the SS Newtonian potential which is used to compute the positions and three-dimensional motions of the Earth-Sun, Moon-Sun and Saturn-Sun SPs. 
Following the lines of previous work \cite{bekenstein2006,bevis2010}, we then use our Newtonian model to compute the QMOND phantom DM distribution within
the low-gravity bubbles surrounding the SPs. 
In Sec. \ref{section4}, we present and discuss numerical solutions to the QMOND potential near the Earth-Sun SP. We further derive the extra tidal stress felt by a spacecraft near the saddle and readdress the question of detecting such signals with the LPF.
Considering the SP between two massive bodies, we also derive semi-analytic solutions for the potential which, together with a comparison to a full numerical treatment, are separately given in an appendix. Finally, we conclude in Sec. \ref{section5}.

\section{Low acceleration bubbles in the solar system}
\label{section2}
\subsection{Quasi-linear MOND}
\label{section2a}
QMOND or (QUMOND) \cite{qmond} is a recently proposed quasi-linear formulation of the original MOND paradigm \cite{Mond1,Mond2,Mond3}. Within this framework, the
gravitational potential $\Phi$ generated by a mass distribution $\rho_{\rm bary}$ is a solution of Poisson's equation for the modified source density
\beq
\Delta\Phi = 4\pi G\hat{\rho} = -\bm{\nabla}\cdot\left\lbrack \nu (g_{N}/a_{0})\mathbf{g}_{N}\right\rbrack ,
\label{eq:m1}
\eeq
where $g_{N}\equiv |\mathbf{g}_{N}|$ is the Newtonian acceleration and the acceleration scale $a_0 \approx 1.2\times10^{-10}$ m/s$^{-2}$ is Milgrom's constant. Here the free function $\nu(y)$ is related to the MOND interpolation function $\mu(x)$ according to $\nu(y) = 1/\mu(x)$ and $y = x\mu(x)$,
where $x=g/a_{0}$ is the MOND total gravity measured in units of $a_0$ \footnote{Note that a real correspondence between these interpolating functions only exists for situations where contributions due to the ``curl'' term vanish, which, for instance, is satisfied in spherically symmetric configurations.}. In the deep-MOND regime, i.e. $g_{N}\ll a_{0}$, we have that $\nu(g_{N}/a_{0}) \approx (g_{N}/a_{0})^{-1/2}$ whereas $\nu\rightarrow 1$ for strong gravitational fields, i.e. $g_{N}\gg a_{0}$, in which case Eq. \eqref{eq:m1} reduces to the usual Poisson equation. For later purposes, it is convenient to express the modified QMOND density $\hat{\rho}$ as the sum of the baryonic contribution $\rho_{\rm bary}$ and a phantom dark matter (PDM) term $\rho_{\rm PDM}$:
\beq
\hat{\rho} = \rho_{\rm bary} + \rho_{\rm PDM}.
\label{eq:m1b}
\eeq
As has been discussed in Ref. \cite{qmond}, QMOND is derivable from an action and thus satisfies all the usual conservation laws. Furthermore, it is important to
note that, as has been shown in Refs. \cite{bimetric, bimetric2}, our Eq. (1) is the genuine nonrelativistic limit of a particular class of bimetric gravity theories
\footnote{Given a particular system in the bimetric framework, this assumes that the influence of matter fields, which are exclusively coupled to the auxiliary metric,
is negligible or that such ``twin'' matter simply does not exist. The QMOND interpolation function is then obtained from the coupling function $\mathcal{M}$, which
appears in the interaction term between the two metrics, through the relation $\nu(y) = 1 + \mathcal{M}^{\prime}(y^{2})$, where $\mathcal{M}^{\prime}\equiv d\mathcal{M}(z)/dz$.}.
Contrary to TeVeS and similar proposed relativistic extensions, these constructions do {\it not} necessarily require a renormalization of the gravitational constant, meaning a difference between the bare value of $G$ (present in the Lagrangian) or $G_{c}$ (the effective value used in cosmology calculations) and the usual Newtonian gravitational constant measured on Earth. In Sec. \ref{section2b}, we will further elaborate on this and its implications for the present analysis.

The practical advantage of the QMOND formulation with respect to classical MOND is the fact that it requires only solutions to linear partial differential equations, with a further nonlinear algebraic step. 
In general, the QMOND density $\hat{\rho}$ can be interpreted as the modified source density that yields a MOND-like potential. 
Using simple algebra, we can express Eq. \eqref{eq:m1} as
\beq
\hat{\rho} = \nu\left(g_{N}/a_{0}\right )\rho_{\rm bary} - \left(4\pi Ga_{0}\right)^{-1}\nu^{\prime}\bm{\nabla} g_N\cdot\mathbf{g}_{N}.
\label{eq:m2}
\eeq
Near a SP (in the absence of baryonic matter), we have $\rho_{\rm bary}=0$ and the above reduces to 
\beq
\hat{\rho} = \rho_{\rm PDM} = -\left(4\pi Ga_0\right )^{-1}\nu^{\prime}\bm{\nabla}g_{N}\cdot\mathbf{g}_{N},
\label{eq:m3}
\eeq
where 
\beq
\nu^{\prime} \equiv \dfrac{d\nu(y)}{dy}
\label{eq:m4}
\eeq
is the derivative of the interpolation function.
The PDM concept is useful to visualize the effects of nonlinearity in MOND using the standard Newtonian picture: For example, it was shown that, at galactic scales,
the PDM may generate an effective DM disk \cite{nipoti2007,zhao2010comp}. Other work demonstrated that placing a galaxy into an external graviational field leads to
an hourglass-shaped distribution of negative PDM density \cite{wu2008}. 
The symmetry axis of this hourglass-shaped distribution  is parallel to the direction of the external field (see, e.g., Fig. $3$ of Ref. \cite{wu2008}). Presuming that DM particles always give rise to a positive dynamical density, this points toward one of the most important differences between (cold) DM and MOND at galactic scales. At solar system scales, the situation is quite similar since, as we show in Sec. IV, an analogous distribution of negative PDM is present near the SPs of the total Newtonian potential in the SS.

\subsection{QMOND bubbles}
\label{section2b}
It has been claimed that a direct test of Newton's laws in the regime of ultra-low accelerations is feasible with our current technological capabilities.
More precisely, it has been pointed out that a spacecraft, closely approaching one of the low-gravity bubbles near the Earth, should be able to detect an
additional tidal stress signal in MOND whose amplitude and shape has been predicted in the framework of TeVeS \cite{bekenstein2006,bevis2010}. Naively, one
would expect these bubbles to be relatively small since MOND effects should become important only for accelerations smaller or approximately equal to $a_{0}$. A simple
Newtonian calculation shows that the semi-major axis of the ellipsoid defined by the condition $g_{N} = a_0$ is $\approx5$ m for the Earth-Sun saddle and $\approx1$ m
for the Moon-Sun saddle. In TeVeS, however, the dynamics of MOND enters through an additional scalar field $\phi$, i.e. $\Phi\approx\Phi_{N} + \phi$, where
$\Phi_N$ is the usual Newtonian potential (see App. \ref{appendix2}). Thus, even if the total potential $\Phi$ is still dominated by its Newtonian
contribution, it is possible to have $\phi$ already in the deep-MOND regime. This peculiarity has been exploited in Ref. \cite{bekenstein2006} to estimate
a characteristic acceleration scale in TeVeS below which full MONDian behaviour of $\phi$ should be triggered, $a_{\rm trig} = (4\pi/k)^{2}a_{0}$, where $k$
is the TeVeS scalar parameter. Dividing $a_{\rm trig}$ by the tidal stress at a generic saddle point then yields a corresponding TeVeS bubble with major axis
\beq
r_{\rm trig} = \frac{a_{\rm trig}}{T},
\label{eq:m5}
\eeq
where $T$ is the Newtonian tidal stress along the symmetry axis in the two-body approximation,
\beq
T = \frac{2 GM}{{(d-r_{\rm sp})}^3}\left (1 + \sqrt{M/m}\right ),
\label{eq:m6}
\eeq 
$r_{\rm sp} \approx d\sqrt{m/M}$ is the distance of the SP from the lighter body (assuming $M>>m$), and $d$ denotes the separation between the two bodies.
For example, the Newtonian tidal stress at the Earth-Sun SP is $T \approx 4\times 10^{-11}$ s$^{-2}$. Fixing $k=0.03$ throughout this work, we have
$a_{\rm trig}\approx 10^{5}a_{0}$ and the above formulae allow one to estimate a bubble size of $766$ km for the Earth-Sun bubble, $280$ km for the
Moon-Sun, and $3.57\times 10^6$ km for the Saturn-Sun bubble.

As for the framework of QMOND and its relativistic bimetric formulation, we assume that the value of $G_{c}$, i.e. the effective gravitational constant
that is relevant to cosmology (e.g. in nucleosynthesis calculations), is identical to both the bare $G$ (entering the theory's Lagrangian) and the value of
$G$ measured on Earth (i.e. $G=G_{c}=G_{\rm bare}$) \cite{milgrom2012novel,milgrom2010cosmological}. In the context of Milgrom's general bimetric MOND gravity
\cite{bimetric,bimetric2}, such a situation explicitly corresponds to setting $\alpha + \beta = 0$ and $\beta =1$. This leaves substantial freedom for choosing
viable interpolation functions $\nu(y)$ (cf. Ref. \cite{magueijo2011}), and also means that our Eq. \eqref{eq:m1} in  Sec. \ref{section2a} is rigorously correct.
Despite a first investigation of Friedmann-Robertson-Walker solutions \cite{Clifton2010}, the full cosmological scenario in the framework of these theories
remains to be worked out completely, and there are presently no constraints on how fast $\nu(y)$ may approach the Newtonian limit. In the rest of the paper, we
therefore take $G=G_{c}=G_{\rm bare}$, bearing in mind that new cosmological constraints on $\nu(y)$ might arise in the future. An important consequence of this
assumption is that a characteristic bubble size akin to Eq. \eqref{eq:m5} does not exist in QMOND. Indeed, the bubble sizes associated with a detectable anomalous
stress signal will sensitively depend on the assumed form of $\nu(y)$, a problem we will address in Sec. \ref{section4}. However, we will nevertheless use
$r_{\rm trig}$ given by Eq. \eqref{eq:m5} to define bubble sizes when addressing the problem of their motions and evolution in Sec. \ref{section3}.

A necessary prerequisite for any realistic performance of the LPF flyby experiment is the precise knowledge of the positions and motions of the
low-acceleration bubbles near the Earth. Neglecting extragalactic gravitational fields (which typically produce accelerations of
$a_{\rm ext}<<10^{-10}$ m s$^{-2}$), the total Newtonian gravitational potential at a point within the SS is due to the contributions of the Sun,
the planets (as well as their satellites) and the gravitational field of the Milky Way. The exquisite knowledge of the relative magnitudes of their
Newtonian potentials allows one a precise determination of the SP positions within the SS and the shape of the isopotential surfaces nearby. An
approximate estimate of perturbations induced by a constant external acceleration field of magnitude $a_p$ on the position of a SP is given by
(see Ref. \cite{bekenstein2006})
\beq
\delta r \approx \frac{2a_{p}}{T}.
\label{eq:m7}
\eeq 
Substituting values for the Earth-Sun SP, we find that the Moon induces a mean perturbation on the two-body position of the SP of the order of
$\delta r \approx 6000$ km while the presence of Jupiter leads to a perturbation of $\delta r \approx 10$ km. However, the gravitational field generated by the Milky Way also plays a role in determining the exact position of the SP. Its gravitational acceleration across the SS can be estimated assuming that the Sun's orbit around the center of the Galaxy is in centrifugal equilibrium. Using this hypothesis, we have ${v^2}/ r \approx 1.9\times 10^{-10}$ m s$^{-2}$ (assuming $v=220$ km s$^{-1}$, $r=2.5 \times 10^{17}$ km), three orders of magnitude less than the gravitational pull exerted by Jupiter. This would translate into a positional shift of the SP of a few cm, and can therefore be neglected.

From these back-of-the-envelope calculations, we conclude that, within the three-body approximation the position of the Earth-Sun SP can be tracked with an accuracy of $\approx$10 km.  A plausible impact parameter for the LPF spacecraft has been estimated as $b\approx 50$ km \cite{bevis2010}. As we will show in the next sections, however, the strength of the MOND tidal stress signal decreases substantially with the ''rapidity'' of the transition between the Newtonian and MOND regimes, which is determined by the slope of $\nu(y)$. As a consequence of this, to test a wider range of interpolation functions, the LPF spacecraft might need to approach the SP more closely than previously thought. This motivates our choice to use a many-body code to calculate the true motion of the SPs to excellent accuracy given by our current knowledge of the SS.   
\begin{table*}
\caption{Table of adopted constants: All values are expressed in SI units.}
\begin{ruledtabular}
\begin{tabular}{c c c c c}
\noalign{\smallskip}
{\bf Name} & {\bf Symbol} & {\bf Value} & {\bf Uncertainty} & {\bf Reference}
\tabularnewline
\noalign{\smallskip}
\hline
\noalign{\smallskip}
Gaussian gravitational constant & $k_{G}$ & 0.01720209895 & - & IAU 1976, \cite{pitjeva2009}
\tabularnewline
Astronomical unit & AU & 149597870700 m & $\pm$3 m & \cite{pitjeva2009}
\tabularnewline
Heliocentric gravitational constant & $GM_{\odot}$ & 1.32712442099$\times10^{20}$ m$^3$s$^{-2}$ & $\pm$1$\times10^{10}$ & \cite{folkner2008}
\tabularnewline
Geocentric gravitational constant & $GM_{\oplus}$ & 3.986004418$\times10^{14}$ m$^3$s$^{-2}$ & $\pm$8$\times10^{5}$ & \cite{ries1992}
\tabularnewline
$M_{\odot}/M_{\rm Mercury}$ & - & 6.0236$\times$10$^{6}$ & $\pm$3$\times10^{2}$ & \cite{anderson1987}
\tabularnewline
$M_{\odot}/M_{\rm Venus}$ & - & 4.08523719$\times$10$^{5}$ & $\pm$8$\times10^{-3}$ & \cite{konopliv1999}
\tabularnewline
$M_{\odot}/M_{\rm Mars}$ & - & 3.09870359$\times$10$^{6}$ & $\pm$2$\times10^{-2}$ & \cite{konopliv2006}
\tabularnewline
$M_{\odot}/M_{\rm Jupiter}$ & - & 1.047348644$\times$10$^{3}$ & $\pm$1.7$\times10^{-5}$ & \cite{jacobson2000}
\tabularnewline
$M_{\odot}/M_{\rm Saturn}$ & - & 3.4979018$\times$10$^{3}$ & $\pm$1$\times10^{-4}$ & \cite{jacobson2006}
\tabularnewline
$M_{\odot}/M_{\rm Uranus}$ & - & 2.290298$\times$10$^{4}$ & $\pm$3$\times10^{-2}$ & \cite{jacobson1992}
\tabularnewline
$M_{\odot}/M_{\rm Neptune}$ & - & 1.941226$\times$10$^{4}$ & $\pm$3$\times10^{-2}$ & \cite{jacobson2009}
\tabularnewline
\noalign{\smallskip}
\end{tabular}
\end{ruledtabular}
\label{table1}
\end{table*}

\section{Motion of the MOND bubbles}
\label{section3}
\subsection{The computational framework}
We have used the symbolic calculus package {\it Mathematica} to build a high accuracy model of the SS's Newtonian potential and to compute the positions of the Earth-Sun, Moon-Sun and Saturn-Sun SPs. In computing the total Newtonian potential across the SS, our model takes into account the planetary potential and the contributions of the most massive natural satellites. For the positions of the planets, we use INPOP08 high-precision ephemeris \cite{fienga2009} and adopt the most recent IAU estimates for the Gaussian gravitational constant, the planet's masses relative to the Sun and their gravitational moments. All adopted values and their corresponding references are listed in Table \ref{table1}. Furthermore, note that the oblateness of the Sun, Jupiter, Saturn and the Earth are taken into account when computing the potential.
Assuming polar coordinates, the Newtonian potential of a planet (centered on the planet with the equatorial plane at $z=0$) is represented by the series
\beq
\phi_{N}(r,z) = -\dfrac{Gm}{r}\left\lbrack 1 - \sum\limits_{n=2}^{\infty}\dfrac{1}{r^n}J_n P_n\left(\dfrac{z}{r}\right )\right\rbrack,
\label{eq:m8}
\eeq 
where $J_n$ the $n$-th gravitational moment of the planet, $r= \sqrt{x^2 + y^2 + z^2}$ and $P_n$ the Legendre polynomial of order $n$.
In our code the series is truncated at $n=2$. As the distance of the other planets from the SPs is very large, for them also the terms
of $O(r^{-2})$ can be neglected. Once all contributions to the Newtonian potential have been added together, our code calculates the positions
of the SPs by numerically solving the equation $\bm{\nabla}\phi_N = 0$ using Newton's tangents method.
\begin{figure}
\includegraphics[angle=0,width=\linewidth]{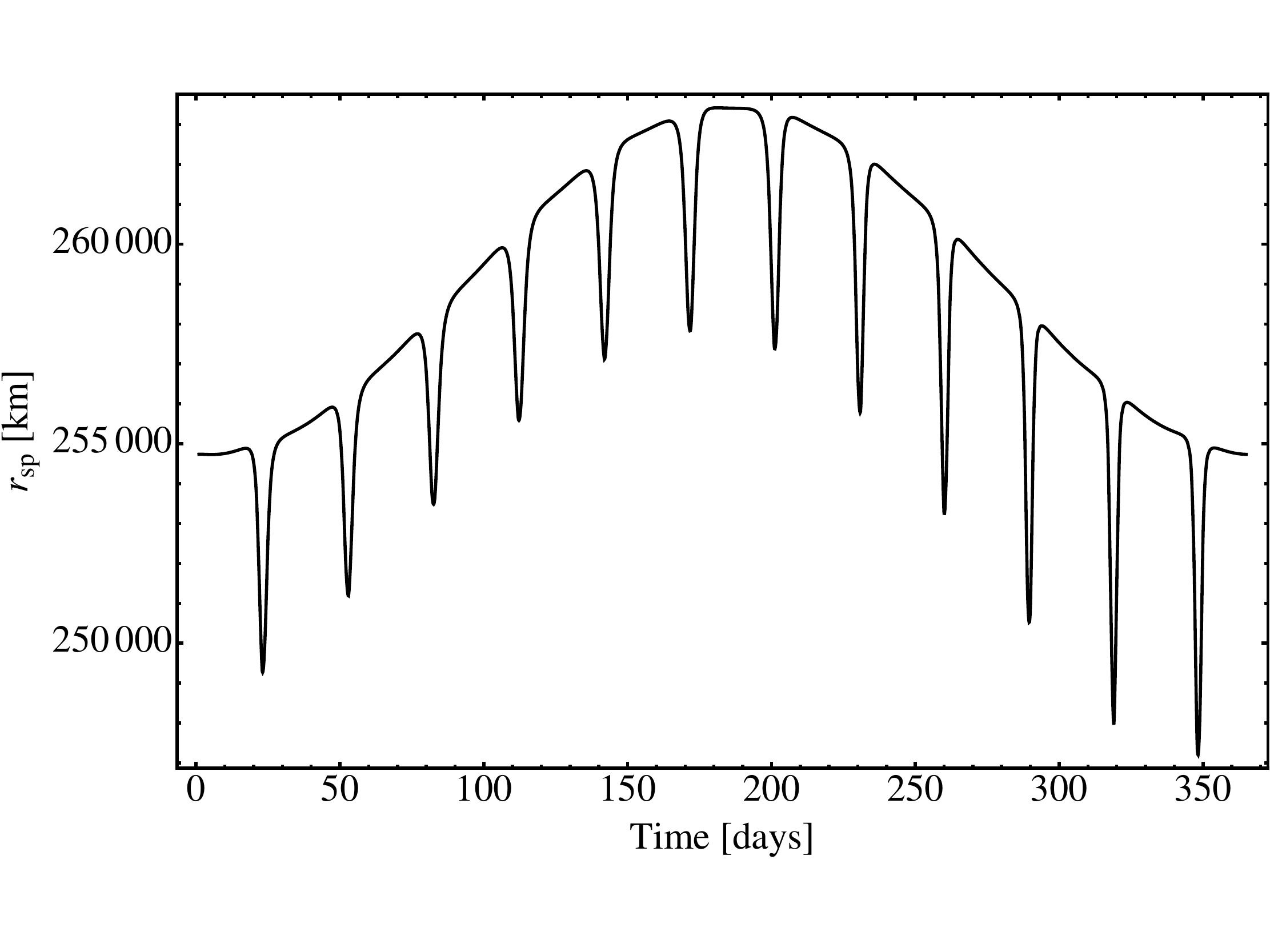}
\caption{Distance of the Earth-Sun SP from the Earth during a year: The gentle yearly variation is due to the eccentricity 
on the Earth's orbit and its amplitude is (measured on the picture) around $\Delta r_{\rm sp} = 8800$ km. The vertical peaks
are due to the moon's perturbation. Their measured amplitudes oscillate between  $4700$ km $< a < $ $9000$ km.}
\label{fig:1}
\end{figure}

\begin{figure}
\centering
\includegraphics[angle=0,width=1\linewidth]{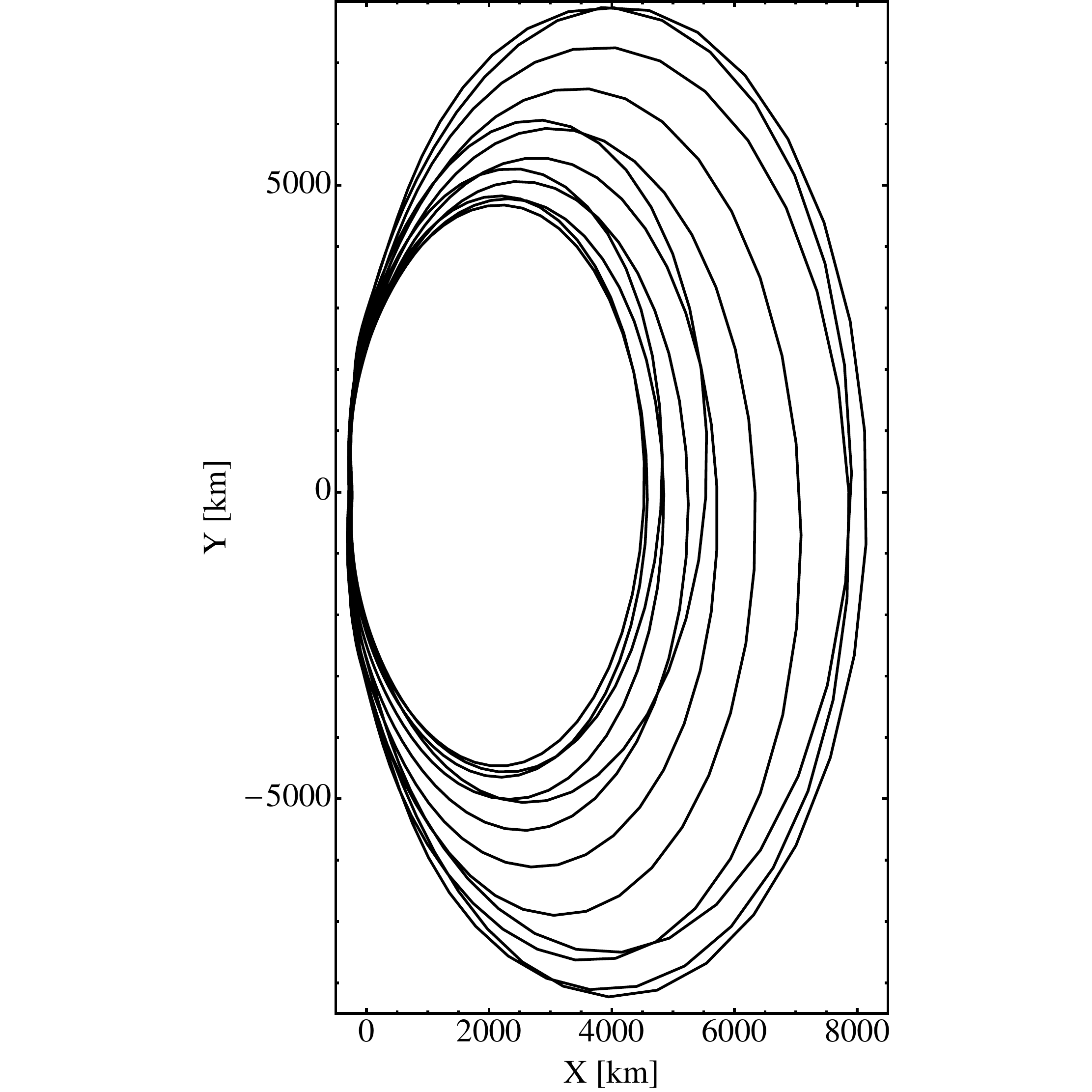}
\caption{Epicycles described by the Earth-Sun SP around its unperturbed position in a year (projected on the plane of the ecliptic; here the $xy$-plane).}
\label{fig:2}
\end{figure}

\subsection{The Earth-Sun SP}
In general, the SS SPs of the total Newtonian potential follow non-inertial trajectories. An intuitive picture of the characteristics of this motion can be obtained by starting with the two-body approximation and considering the effects of a perturber later. Consider two bodies of masses $M$ and $m$, respectively, with $M\gg m$. Choose a corotating Cartesian coordinate frame centered on the orbiting object with the $x$-axis pointing toward the heavier body and call the $xy$-plane the ecliptic plane. In this reference frame, the SP is stationary if the orbit of the lightest body is circular, and its distance from
the center of the orbit is fixed at $r_{\rm sp}$. If the orbit is instead elliptical, then a periodic oscillation of the SP along the line joining the two bodies arises.
This oscillation has the same period as the orbiting body and an amplitude that is given by
\beq
\Delta r_{\rm sp} = \Delta d \sqrt{\frac{m}{M}},
\label{eq:m9}
\eeq 
where $\Delta d$ is the difference between the apoapsis and periapsis of the orbiting body. For the Earth-Sun SP, the above equation gives
$\Delta r_{\rm sp}\approx 8666$ km. If a perturber is added to this two-body system (for example, the Moon in the case of the Earth-Sun SP), then, besides this
gentle yearly variation due to the orbital eccentricity, the SP will also describe an epicycle around its unperturbed position. The semi-major axis of this epycicle is given by Eq. \eqref{eq:m7}. The minor axis of this epicycle is half of its major axis and it is parallel to the line joining the two major bodies. The period of this epicycle is given by the period of the perturber (if it orbits around the lighter body).
For the Earth-Sun SP, the perturber is the Moon. The mean semi-major axis of the Earth-Sun bubble epicycle due to the Moon's perturbation is $a \approx 6000$ km and its period equals the synodic period of the Moon. Since the Moon orbits on a plane inclined with respect to the ecliptic plane, the SP has also a component of motion in the direction perpendicular to the ecliptic plane. Therefore, the yearly motion of the Earth-Sun SP is a complicated three-dimensional trajectory, a composition of the
different dynamical effects mentioned above. Using our code, we have produced several plots to illustrate these complicated motions. In Fig. \ref{fig:1}, we show the distance of the Earth-Sun SP from the Earth during a year. 

\begin{figure}
\centering
\includegraphics[angle=0,width=1\linewidth]{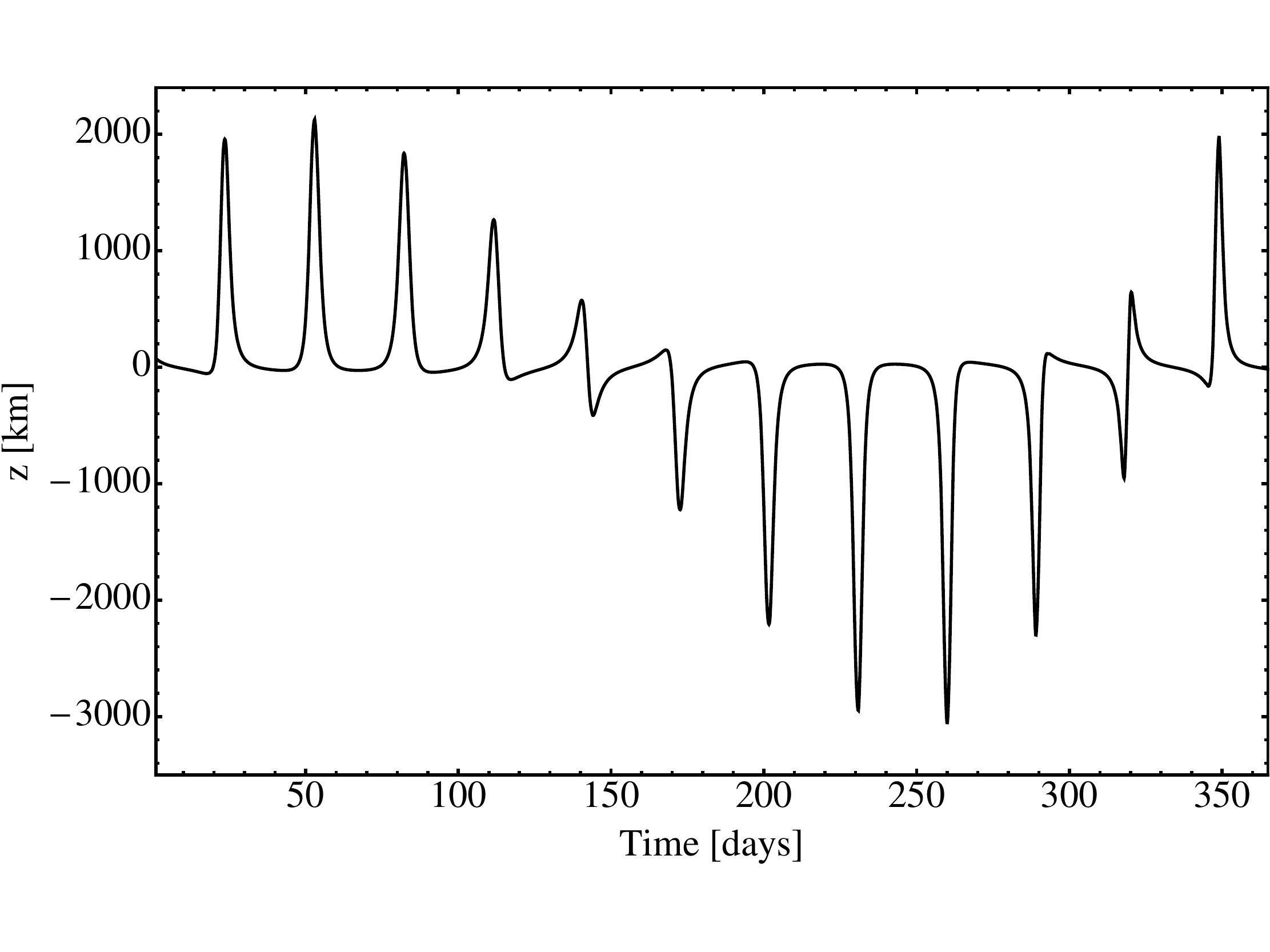}
\caption{Distance of the Earth-Sun SP from the ecliptic plane: This component of the motion is due to the inclination of the Moon's orbital plane with respect to the ecliptic. The peak-to-valley amplitude of this yearly oscillation is around $5000$ km.}
\label{fig:3}
\end{figure}

The gentle yearly variation of the distance due to the Earth's eccentric orbit is roughly $\Delta r_{\rm sp} = 8800$ km, very close to the value calculated with the help
of Eq. \eqref{eq:m9}. The vertical spikes caused by the Moon's presence have amplitudes varying between $4700$ km$<a<$9000 km, with a mean value of around $6850$ km.
This is compatible with the amplitude calculated from Eq. \eqref{eq:m7}. In Fig.\ref{fig:2}, we show the epicycles (projected onto the ecliptic plane) described by the Earth-Sun SP around its unperturbed (two-body) position. In this plot, the Earth lies on the right-hand side. At new moon, the SP shifts toward the Earth, whereas it shifts toward the Sun at full moon. The epicycle's amplitude is higher if new moon occurs at perigee and lower if at apogee. The yearly motion of the Earth-Sun SP in the direction perpendicular to the ecliptic plane is illustrated in Fig.\ref{fig:3}. 

\begin{figure}
\centering
\includegraphics[angle=0,width=1\linewidth]{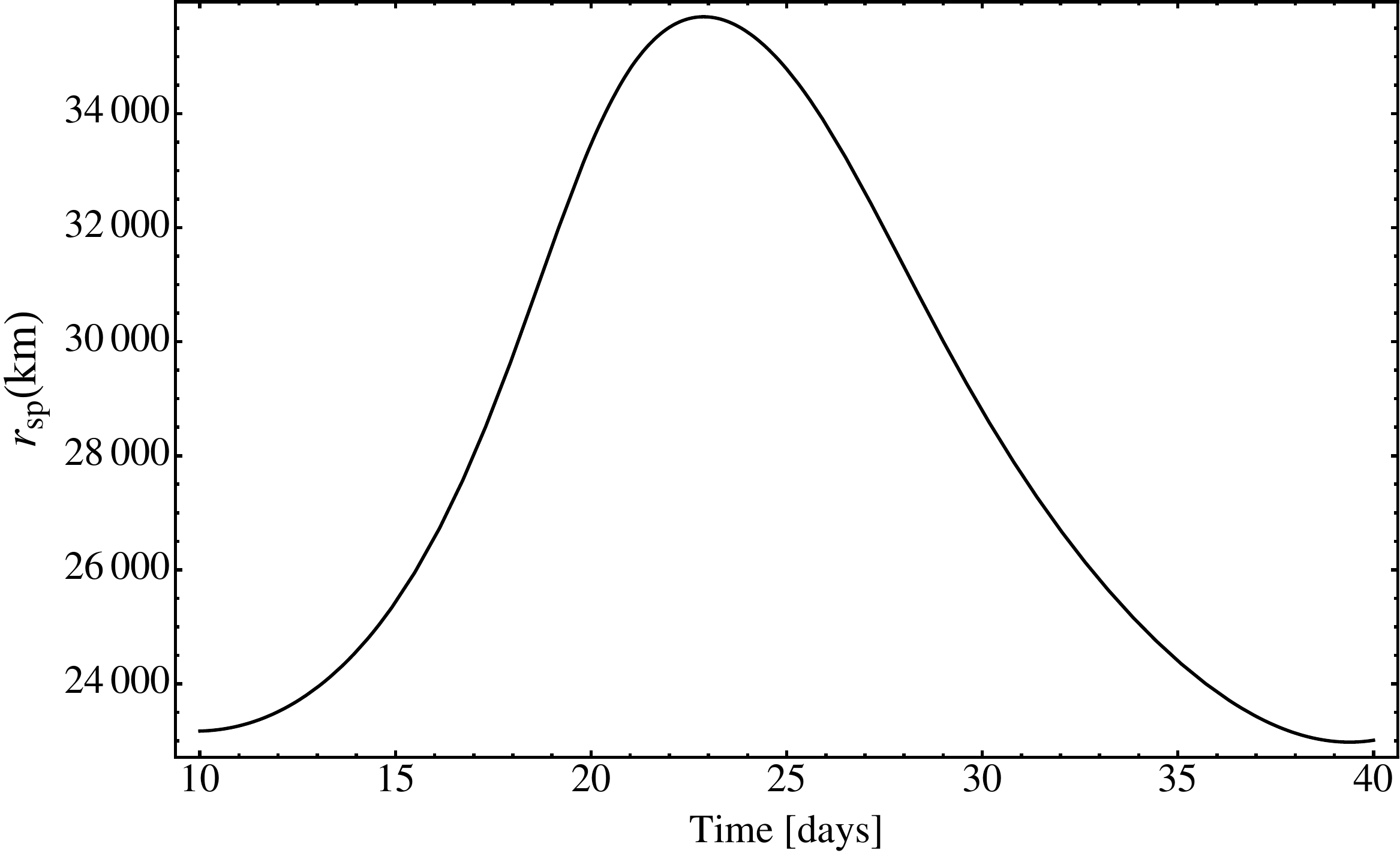}
\caption{Distance of the Moon-Sun SP from the Moon during a sidereal month: The asymmetry of the bell-like curve is due to the inclination of the Moon's orbit.}
\label{fig:4}
\end{figure} 

\subsection{The Moon-Sun SP}
\begin{figure}
\centering
\includegraphics[angle=0,width=1\linewidth]{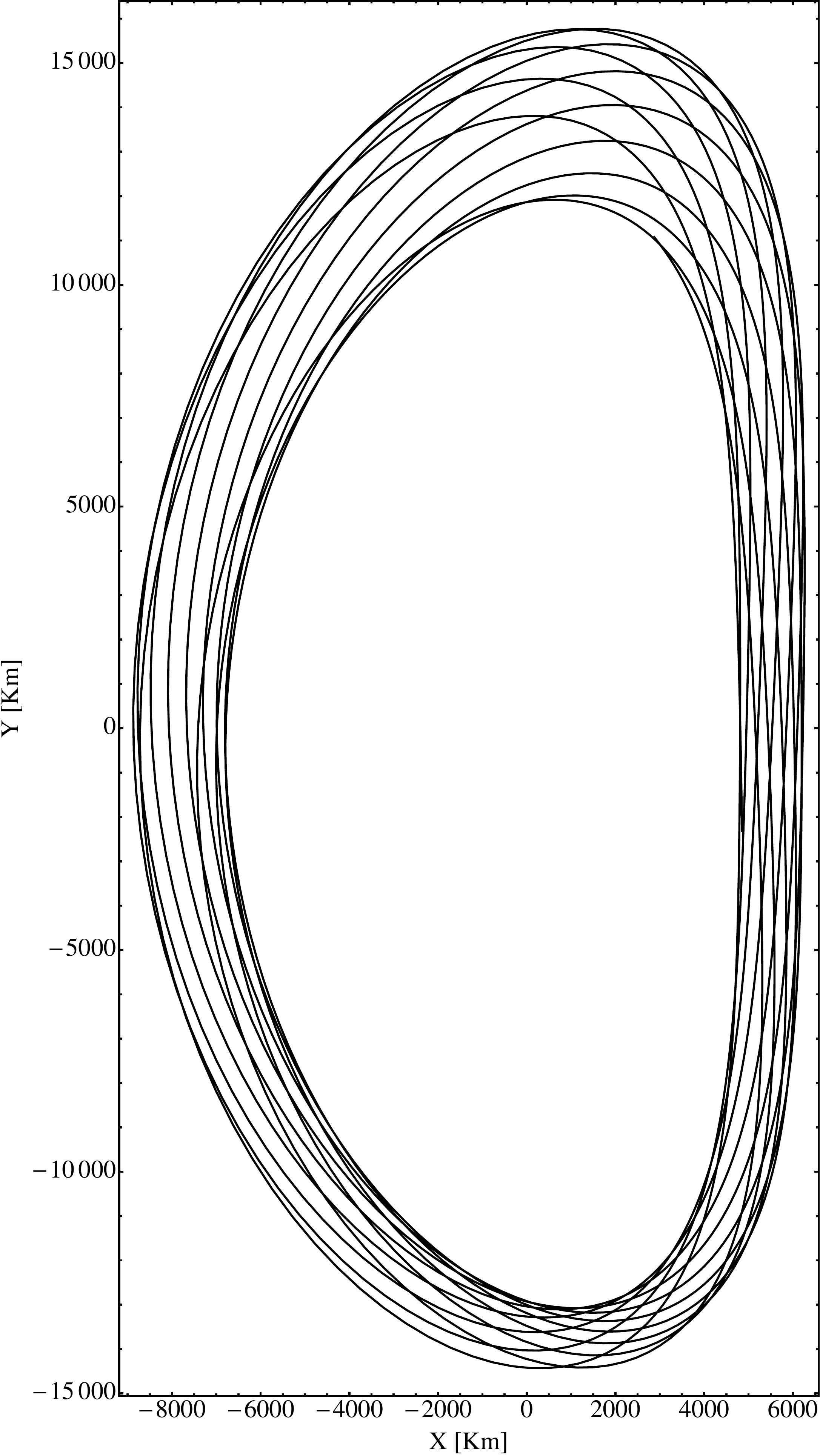}
\caption{Epycicles described by the Moon-Sun SP around its unperturbed position during a year: Here the Moon lies on the left-hand side.}
\label{fig:5}
\end{figure}

\begin{figure}
\centering
\includegraphics[angle=0,width=1\linewidth]{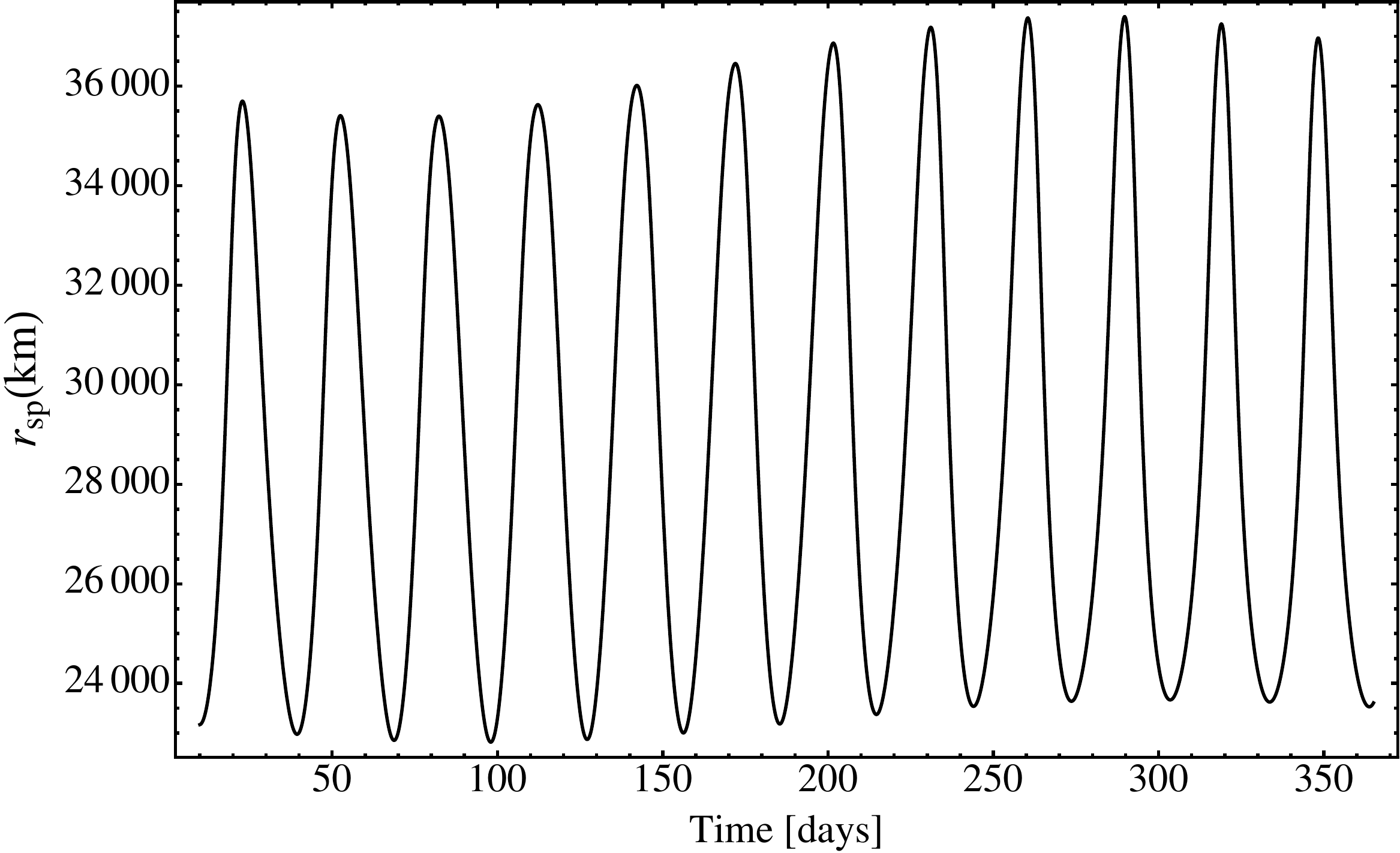}
\caption{Distance of the Moon-Sun SP from the Moon over a year.}
\label{fig:6}
\end{figure}

Another SP close enough to the Earth for direct testing is the Moon-Sun saddle \cite{bevis2010}, sometimes also referred to as the ``Earth-Moon'' SP \cite{bekenstein2006}. However, since the gravitational pull of the Sun on the Moon is higher than that exerted by the Earth, it is better to look at it as a
Moon-Sun SP heavily perturbed by the Earth \cite{bevis2010}. The magnitude of the perturbation induced by the Earth on the position of the bubble is $\Delta r_{\rm sp}\approx 13000$ km, where we have again used Eq. \eqref{eq:m7}. In Fig. \ref{fig:4}, we plot the Moon-Sun SP distance from the Moon for a whole sidereal period. The epicycle's of the SP around its unperturbed position is shown in Fig. \ref{fig:5}. The measured variation in distance during a sidereal month is $\Delta r_{\rm sp}\approx 12550$ km which agrees well with the calculated value. The SP reaches its minimum distance from the Moon at full moon (when it lies between the Earth and the Moon), and its maximum distance during new moon (when it lies between the Moon and the Sun). 
The inclination of the Moon's orbit with respect to the ecliptic changes the Moon-Sun saddle's apoapsis. 
This gentle variation can be seen in Fig. \ref{fig:6}, where we show the SP distance from the Moon's center for a whole year. The peak-to-valley variation in the apoapsis has an amplitude of approximately $2000$ km.

\begin{figure}
\centering
\includegraphics[angle=0,width=1\linewidth]{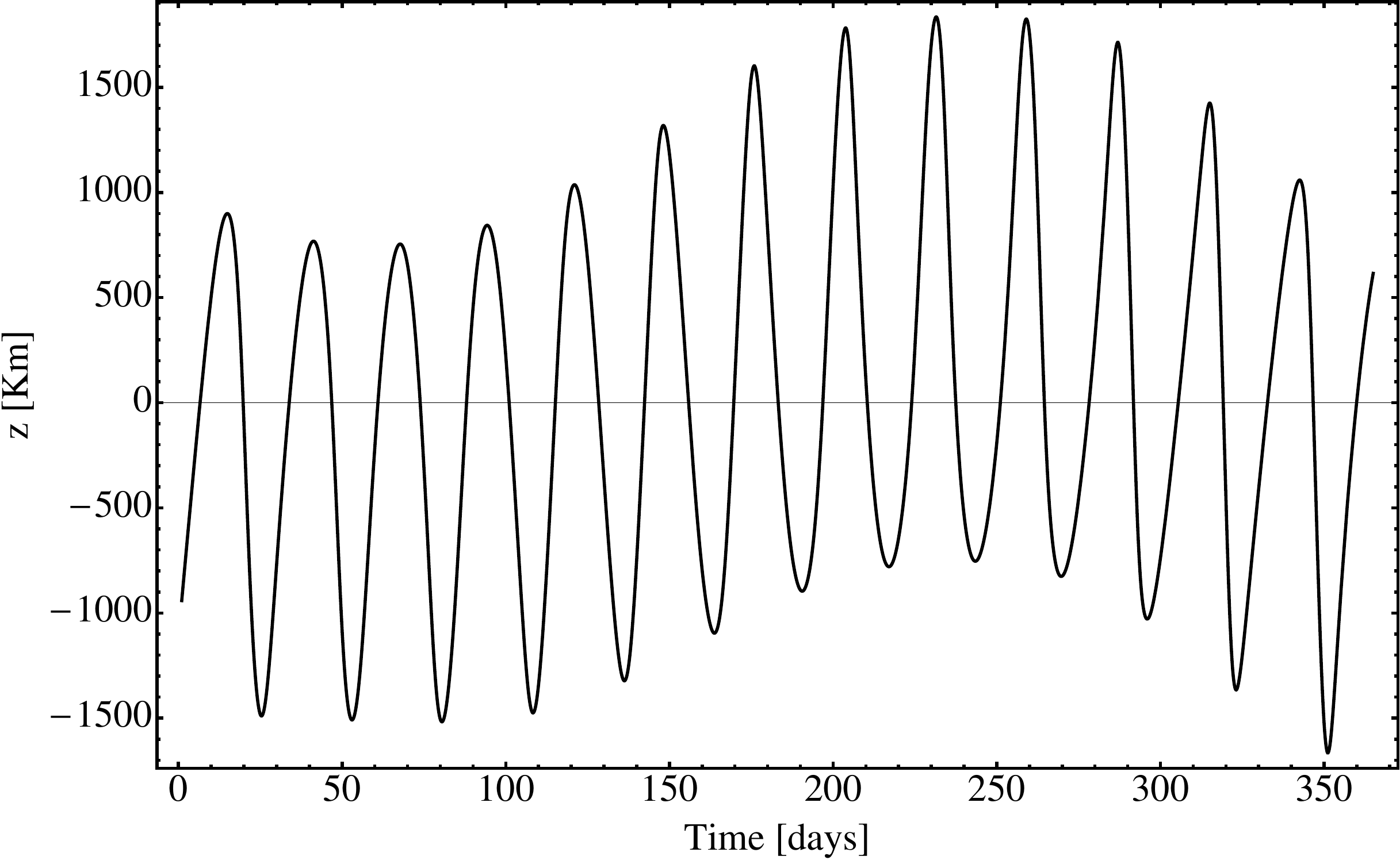}
\caption{Distance of the Moon-Sun SP from the plane of the ecliptic: This component of the motion is due to the inclination of the Moon's orbital plane with respect to
the ecliptic. The peak-to-valley amplitude of this yearly oscillation is around $3000$ km.}
\label{fig:7}
\end{figure}

A question raised in Ref. \cite{bekenstein2006} was whether it may happen (considering the large magnitudes of perturbations) that the Moon-Sun SP crosses the Moon's surface during its motion. From our code, we find a minimum distance of the saddle from the Moon's center of $22512$ km. Since the Moon's radius is $1738$ km, we conclude that the SP never crosses the Moon's surface. In Fig. \ref{fig:7}, we show the motion of the SP in the direction perpendicular to the ecliptic (due to the inclination of the Moon's orbit). For the Moon-Sun saddle, the amplitude of this motion component is lower ($3000$ km compared to $5000$ km) than in the case of the Earth-Sun saddle. This can be explained as follows: Whereas the Moon (with his inclined orbit) tends to push the Earth-Sun saddle away from the ecliptic plane, the combined pull of the Sun and the Earth tends to push the Moon-Sun SP toward the ecliptic plane. 

\subsection{The Saturn-Sun SP}
\begin{figure}
\centering
\includegraphics[angle=0,width=1\linewidth]{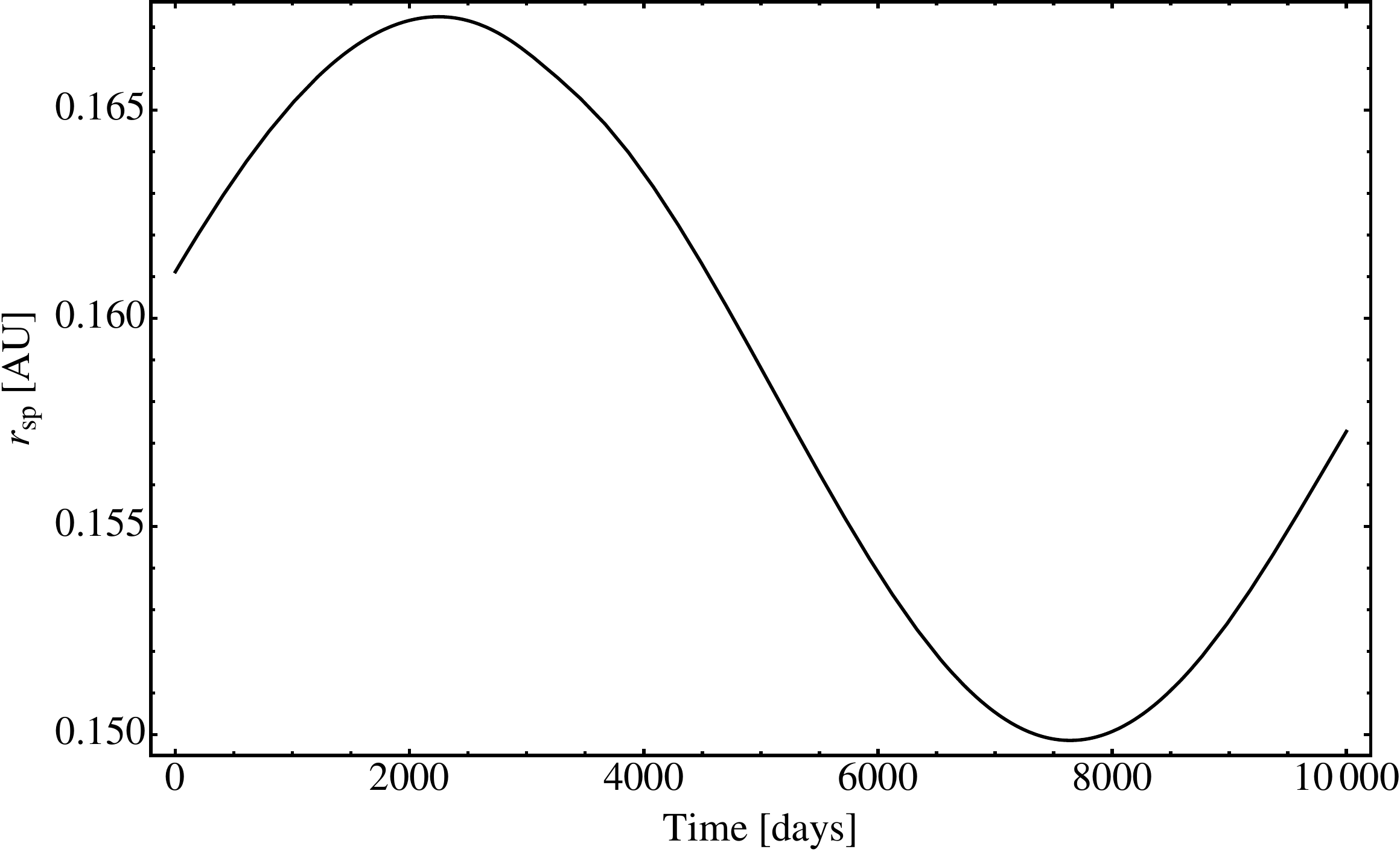}
\caption{Distance of the Saturn-Sun saddle point from saturn during a Saturn's year: The amplitude of the mainly due to the eccentricity of Saturn's orbit is around $0.017$ AU.}
\label{fig:8}
\end{figure}

The Saturn-Sun low-acceleration bubble is one of the largest in the SS. The semi-major axis of the MOND bubble is $3.57\times 10^6$ km $\approx 0.02$ AU. The minimum distance between Jupiter and Saturn is around $5.3\times 10^{8}$ km. When this minimum is reached, Jupiter exerts a gravitational pull of $a_{p}\approx 4.4 \times 10^{-7}$m s$^{-2}$ on the Saturn-Sun SP. The Newtonian two-body tidal stress at the Saturn-Sun SP calculated from Eq. \eqref{eq:m6} is $T\approx 5.9\times 10^{-15}$ s$^{-2}$. Using Eq. \eqref{eq:m7}, we may therefore calculate the peak-to-valley amplitude of this perturbation which turns out to be around $1.5\times 10^5$ km. This is much lower than the variation in distance due to the orbital eccentricity which is approximately $2.54\times 10^6$ km. In Fig. \ref{fig:8}, we show the distance of the SP from Saturn during a Saturnian year.

\begin{figure}
\centering
\includegraphics[angle=0,width=1\linewidth]{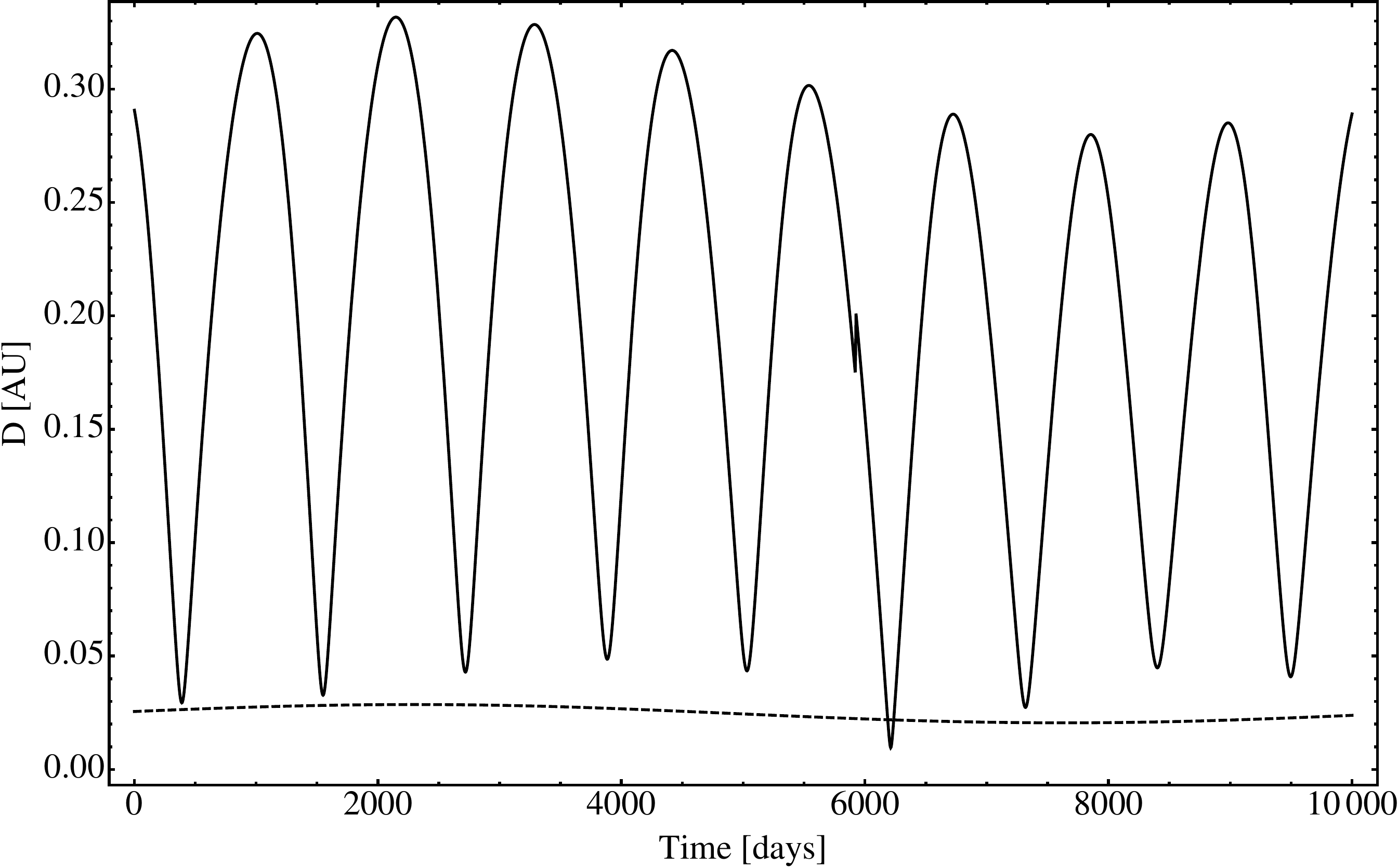}
\caption{Distance of the Saturn's moon Fenrir from the Saturn-Sun SP during a Saturnian year: The horizontal dashed line shows the size of the MOND bubble according to Eq. \eqref{eq:m5}. Note that $6212$ days after the 2012 Jan 1, the Moon gets very close to the center of the bubble at a distance of only $r_{\rm min}\approx$ 0.437$r_{\rm trig} \approx 1.43\times 10^{6}$ km.}
\label{fig:9}
\end{figure}

Considering the size of the Saturn-Sun MOND bubble, there is a considerable chance that minor bodies or some of Saturn's outer satellites periodically intersect it. Indeed, we have found that a subsample of the Norse group \cite{saturn,saturn2}, consisting of Surtur,
Ymir, Fenrir and Loge, periodically go through the Saturn-Sun MOND bubble. 
These small satellites are characterized by retrograde motion. The semi-major axes of their orbits vary from $2.24\times 10^7$ km (Fenrir) to $2.30\times 10^7$ km (Loge). Their orbits are periodically intersected by the MOND bubble whose distance from Saturn varies between $2.24\times 10^7$ km to $2.5\times 10^7$ km (see Fig. \ref{fig:8}). Furthermore, one finds that the satellites' inclination with respect to the ecliptic is low and varies from $3^{\circ}$ (Surtur) to $16^{\circ}$ (Fenrir).
Since the MOND bubble stays always close to the ecliptic plane, close encounters with these objects are generally favored.
An example of such an encounter between Fenrir and the Saturn-Sun bubble is shown in Fig. 9. For all of these satellites,
we list their dates of encounter with the Saturn-Sun bubble, their minimum distances from the SP as well as their crossing times in
Table \ref{table2}. 

So far, the found subset of Saturn's satellites comprises the only known objects in the nearby SS that periodically enter the MOND regime. Even though the MONDian influence on the satellites' trajectories is likely to be very small, it cannot be ruled out that the induced perturbations at each encounter may accumulate to a potentially observable effect. Regarding a detailed analysis of the MONDian impact on the path of these satellites, we refer to a forthcoming paper. 

\begin{table}
\caption{Norse's satellites minimum distances to the Saturn-Sun MOND bubble.}
\begin{ruledtabular}
\begin{tabular}{c c c c c}
\noalign{\smallskip}
Satellite & $d_{\rm min}$ & $d_{\rm min}/r_{\rm trig}$ & Date & Crossing time
\tabularnewline
 & $\left (10^{6}{\rm km}\right )$ &  & (dd/mm/yyyy) & (days)
\tabularnewline
\noalign{\smallskip}
\hline
\noalign{\smallskip}
Surtur & 2.00 & 0.48 & 24/06/2015 & 64
\tabularnewline
Ymir & 0.65 & 0.16 & 09/02/2013 & 68
\tabularnewline
Fenrir & 1.43 & 0.43 & 03/01/2029 & 49
\tabularnewline
Loge & 0.60 & 0.17 & 16/03/2026 & 58
\tabularnewline
\noalign{\smallskip}
\end{tabular}
\end{ruledtabular}
\label{table2}
\end{table}

\subsection{Error estimates}
\begin{figure}
\centering
\includegraphics[angle=0,width=1\linewidth]{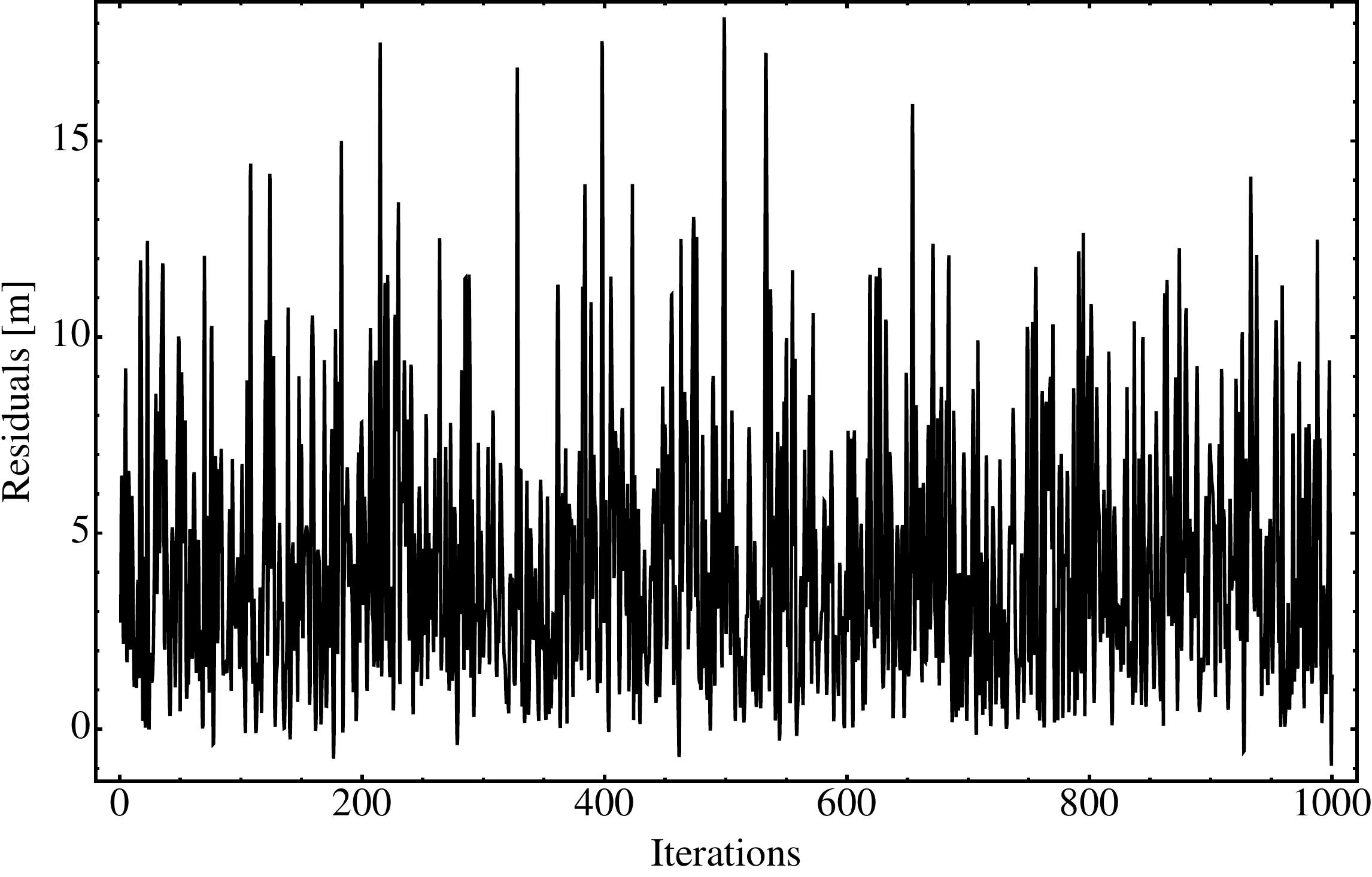}
\caption{Monte-Carlo simulation output: This plot shows the distance of the perturbed Earth-Sun SP from its unperturbed position. The mean value of this data series is a good estimator of our final uncertainty on the position of the Earth-Sun SP.}
\label{fig:10}
\end{figure}

The most important sources of error on the computed positions of the SPs are the uncertainties on the planets' positions and masses. 
We have used a Monte-Carlo method to estimate uncertainties in the positions of the SPs. The masses of the planets are perturbed using random values drawn from
normal distributions with means at the quoted parameter values, and standard deviations matching those listed in Table I. For the error on the planets' position
we use the uncertainties in the INPOP08 ephemeris \cite{fienga2009} which we assume to be roughly $\approx$ 3 m for the Earth's position, a few cm for the Moon's
position, $\approx$ 1000 km for the giant planets and $\approx$ 10 km for the inner planets.
With this method, we obtain an absolute error on the position of the Earth-Sun SP of $\Delta_{\rm E-S} \approx$ 4 m (see Fig. \ref{fig:10}). 
This quantity is close to the error on the two-body position of the Earth-Sun SP $\Delta_{\rm E-S}^{\rm (2-body)} \approx$ 5 m which we obtained by propagating
the errors quoted in Table I to the quantity $r_{\rm sp} \approx d\sqrt{m/M}$ representing the approximate distance of the Earth-Sun SP from the Earth-Moon system
barycenter. Our results tell us that the major sources of error on the position of the Earth-Sun SP are the uncertainties on the $GM_\oplus /GM_\odot$ ratio and
on the AU. The contribution of the  other planets to the final error sum up to an amplitude of about $1$ m. For the Saturn-Sun SP we obtain, in the same way,
an uncertainty of $\Delta_{\rm Sat-S} \approx 2$ km.

\section{QMOND potential inside the bubble}
\label{section4} 
\subsection{Phantom dark matter density near the SP}
\begin{figure}
\centering
\includegraphics[angle=0,width=1\linewidth]{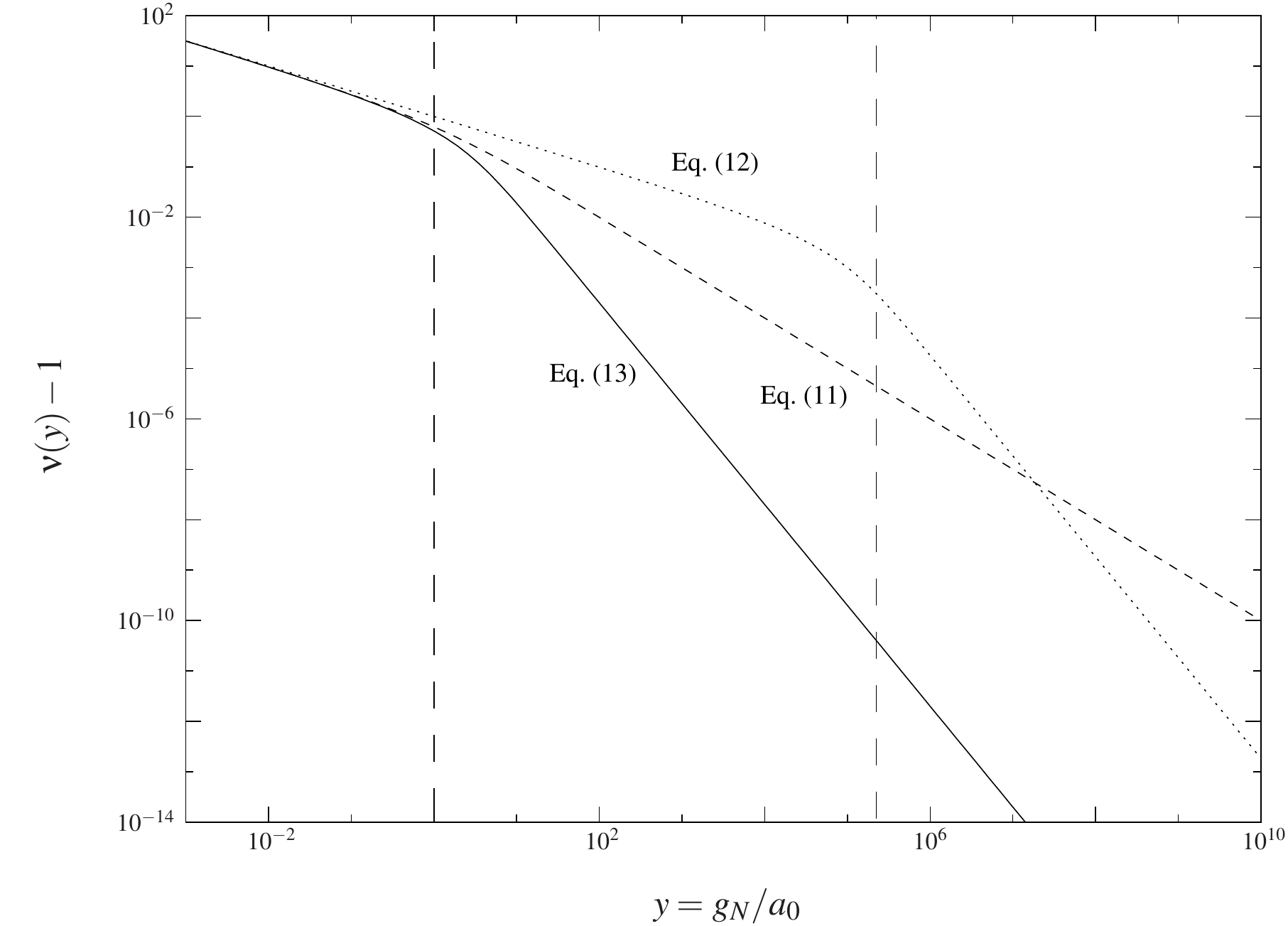}
\caption{Interpolation functions $\nu$ as a function of $y=g_{N}$/$a_0$: Solid, dotted and dashed lines correspond to Eq. \eqref{eq:m11}, Eq. \eqref{eq:m10}
and Eq. \eqref{eq:m12}, respectively. Vertical dashed lines show the characteristic scales $y=1$ and $y=(4\pi/k)^{2}$ for $k=0.03$. The Bekenstein-like
interpolation function Eq. \eqref{eq:m11} gives rise to a much slower transition between the two dynamical regimes.}
\label{fig:11}
\end{figure}

As can be seen from Eq. \eqref{eq:m3}, the PDM density inside a MOND bubble is entirely determined by the Newtonian potential and the choice of
a specific interpolation function. Using our model of the SS gravitational potential, we have calculated the PDM density distribution around the Earth-Sun, Moon-Sun and Saturn-Sun SPs for different choices of $\nu(y)$. In particular, we have used the QMOND analog of the so-called simple interpolation function, $\mu(x)={x/(x+1)}$, which takes the form
\beq
\nu(y) = \frac{1}{2} + \sqrt{\frac{1}{y} + \frac{1}{4}}.
\label{eq:m10}
\eeq
This functional form has been shown to provide a good description of extragalactic phenomenology and allows one, for instance, to fit the observed
rotation curves of a wide range of spiral galaxies very well \cite{famaey2005}. 
In addition, we have considered the interpolation function
\beq
\nu(y)=1-\frac{k}{4\pi}+\left\lbrack\frac{1}{y^{2}} + \left(\frac{k}{4\pi}\right)^4\right\rbrack^{1/4},
\label{eq:m11}
\eeq
which is the QMOND analog of the function used in Refs. \cite{bekenstein2006,bevis2010} and exhibits a behavior close to that of the functional
form originally proposed by Bekenstein in the context of TeVeS \cite{bekenstein2004}.
Finally, we have also chosen an interpolation function similar to Eq. \eqref{eq:m10}, but giving rise to a ''faster'' transition between Newtonian
dynamics and MOND:
\beq
\nu(y) = \frac{1}{2} + \left (\frac{1}{y^{2}} + \frac{1}{16}\right )^{1/4}.
\label{eq:m12}
\eeq
This choice is motivated by recent claims that interpolation functions like Eq. \eqref{eq:m10}, which still result in a relatively ''slow'' transition between
the two dynamical regimes, could give rise to an anomalously high secular precession of the outer planet's perihelion, incompatible with published residuals \cite{blanchet2010, zhao2006}. 
In Fig. \ref{fig:11}, we plot the three interpolation functions as a function of the gravitational field strength expressed in units of $a_0$. As
can be seen, the function defined by Eq. \eqref{eq:m11} approaches the value $\nu(y)=1$ (corresponding to the Newtonian regime) much more slowly than the other ones.
Considering their performance on galactic scales, we briefly discuss all three interpolating functions in App. \ref{appendixrot} where it is also demonstrated that
the behavior of Eq. \eqref{eq:m12} at these scales is very similar to that of the simple $\nu$-function Eq. \eqref{eq:m10}.

\begin{figure}
\centering
\includegraphics[angle=0,width=1\linewidth]{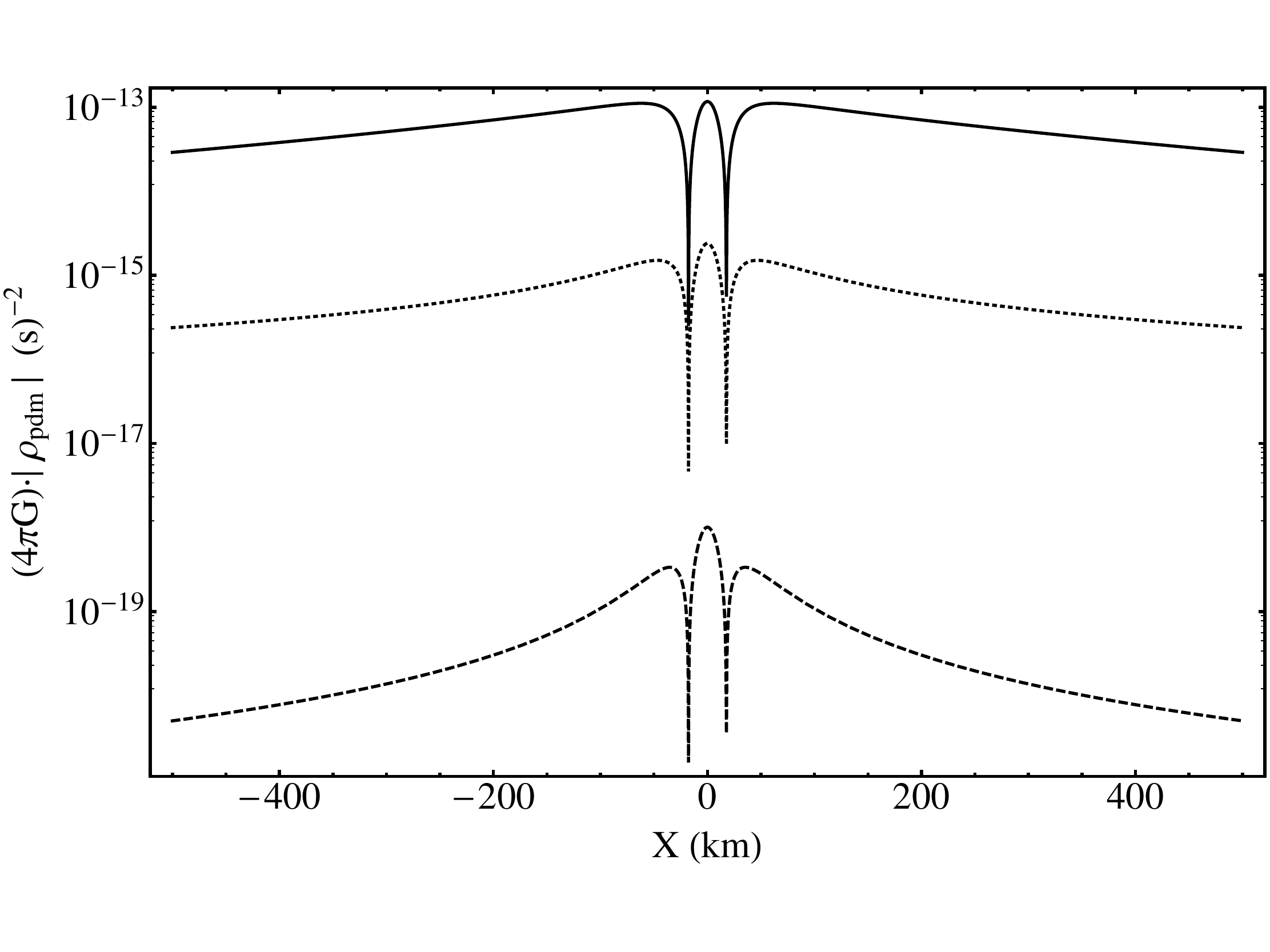}
\caption{Logarithmic plot of the rescaled PDM density's absolute value $4\pi G\rho_{\rm PDM}$ at a cut of $y=50$ km near the Earth-Sun SP: The line styles are defined as in Fig. \ref{fig:11}. Note that the PDM density decreases by several orders of magnitude for the $\nu$-functions with increased slope.}
\label{fig:12}
\end{figure} 

\begin{figure}
\centering
\includegraphics[angle=0,width=1\linewidth]{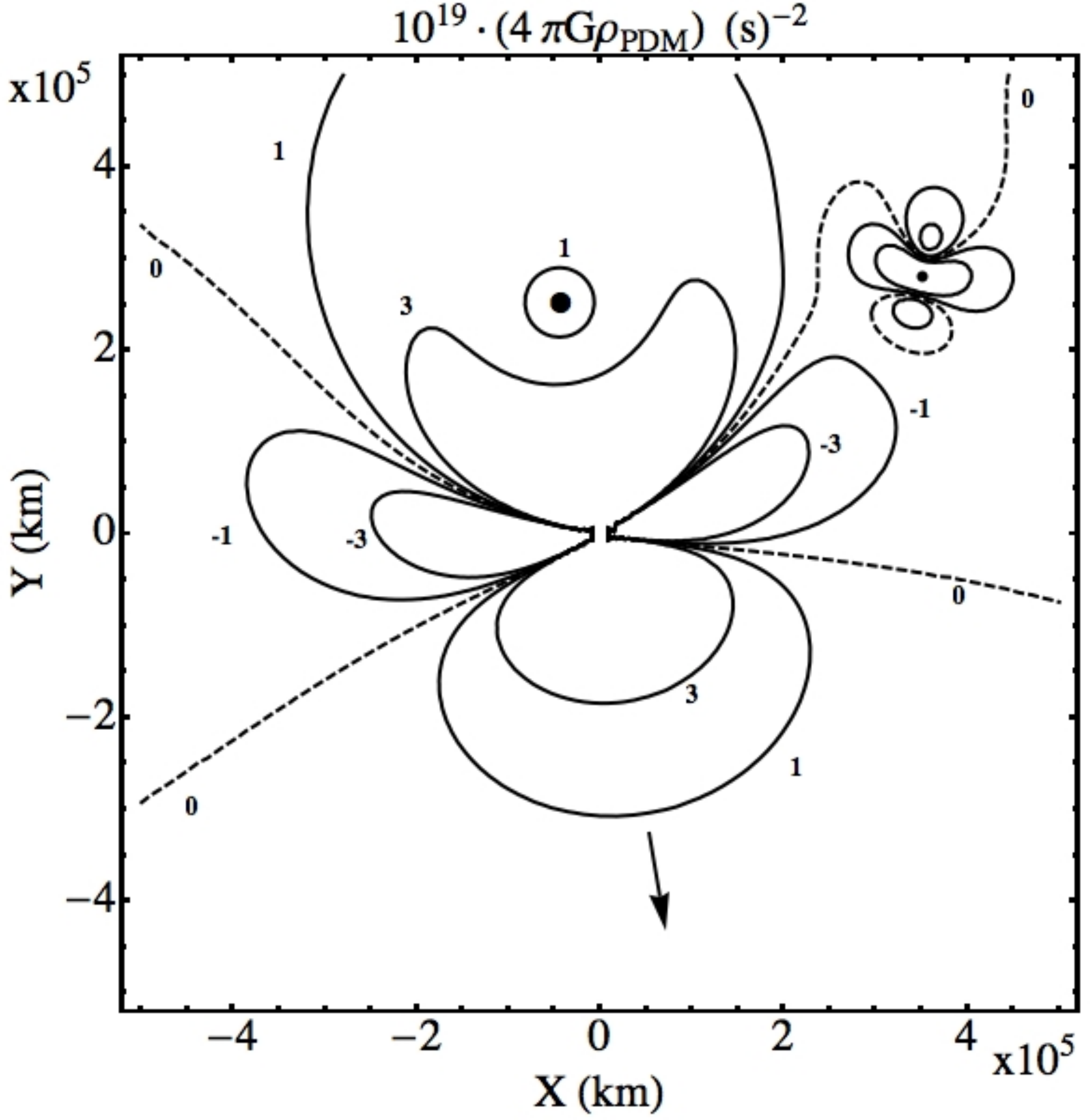}
\caption{PDM isodensity contours inside in a box of $5\times 10^5$ km centered on the Earth-Sun SP along the $z=0$ plane and assuming $\nu$ given
by Eq. \eqref{eq:m11}: The positions of the Earth and the Moon-Sun SP are marked with a large and a small dot, respectively. The arrow points towards the Sun and the dashed line indicates the $\rho_{\rm PDM}=0$ contour. The PDM density is negative within a conic region (deformed by the Moon's presence), with the symmetry axis perpendicular to the Earth-Sun direction, and positive elsewhere. The Earth is embedded into a halo of positive
PDM. Moreover, the PDM density gradient is bigger near the Moon-Sun SP.}
\label{fig:13}
\end{figure}

As remarked earlier, the PDM density depends on $\nu^{\prime}$, which basically tells us that the PDM density within the MOND bubble is sensitive
to the ``transition'' speed between the dynamical regimes. In fact, choosing an interpolation function with a fast enough transition, it is possible
to suppress non-Newtonian effects for accelerations above $a_{0}$. The PDM density near the SP is proportional the trace
of the stress tensor, $4\pi G \rho_{\rm PDM} = {\rm tr}(S_{ij})$, where $S_{ij} = \partial_{i}\partial_{j}\Phi$, and provides a good order-of-magnitude
estimator for the maximum MOND tidal stress signal within a certain region - if one excludes values at special points such as the SPs themselves.

The above dependency on the slope of the interpolation function can be seen in Fig.~\ref{fig:12} where we plot the PDM density at cuts of $y=50$ km near the Earth-Sun SP. It is evident that the PDM density changes by several orders of magnitude between the different choices of $\nu(y)$. We therefore expect that at the same distance (or passing trajectory) from the SP, the corresponding tidal stress signal will be much smaller for an interpolation function given by Eq.~\eqref{eq:m10} than for Eq.~\eqref{eq:m11}. We will confirm this preliminary prediction by using numerical solutions of the QMOND potential near the Earth-Sun SP below.
In Fig. \ref{fig:13}, we illustrate the PDM density distribution calculated with the help of Eq.~\eqref{eq:m11} and rescaled by a factor of $4\times 10^{19}\pi G$ within a box of $10^{5}$ km edge length centered on the Earth-Sun SP. As expected, the PDM density distribution around the SP exhibits cylindrical symmetry, with the symmetry axis along the line joining the planet with the Sun. In the case of the Earth-Sun SP, this symmetry is perturbed by the presence of the Moon.

\subsection{Multi-grid FFT solver}
Since the QMOND PDM decreases quickly toward zero when moving away from the SP, the boundary value problem given in
Eq.\eqref{eq:m1} may be solved by means of efficient numerical methods based on the Fast
Fourier Transform (FFT). To achieve this, we have implemented a modified version of the Poisson solver described
in Ref. \cite{feix2008} which now incorporates a multi-grid scheme to obtain a higher resolution near the SP.

The basic idea is the following: Beginning with a large physical box and a coarse sampling of the underlying density
distribution on an equidistant grid, we solve for the potential $\Phi$ assuming zero boundary conditions - since we
are only interested in the purely non-Newtonian contribution $\Phi_{QM} = \Phi - \Phi_{N}$. Keeping the same number
of grid points per dimension $N$, we then move to a grid of smaller physical size such that the previous solution remains
sufficiently accurate at the new boundaries. A first correction to the potential is then found by considering the solution
for the correction density
field, which is obtained as the difference between the true PDM density and the one calculated by interpolating the
density field on the coarse grid. For this step, we again assume zero boundary conditions and use a cubic spline for the
interpolation. Finally, the above procedure is successively repeated until the desired resolution in the central parts is
reached, and adding the individual results yields the full solution with the appropriate boundary conditions. Apart from
the usual sanity checks (see, e.g., Ref. \cite{feix2008}), the method was applied to several input densities and the corresponding
results were directly compared to those obtained with the original single-grid solver. For our purposes, we have found that the 
relative differences between these results are basically negligible, typically on the order of $\lesssim 10^{-4}$. In App.
\ref{appendix1}, we also present semi-analytic solutions (both for the quasi-Newtonian and the deep-MOND regime) which are used
for comparison to the full numerical results.

In the case of the Earth-Sun saddle, for instance, we set $N=448$ and use three physical cubic boxes with sides
of $20000$, $5000$ and $1250$ km, respectively. This yields a maximum resolution of approximately $\Delta x_{\rm max} = 2.8$ km
in the saddle region. Note that all grids are chosen such that the SP is located at $(0,\Delta x_{\rm max}/2,0)$ to
avoid any numerical difficulties arising from the divergence of the PDM density at the exact SP. Assuming these parameters,
the numerical solution is found to be reliable for distances $\gtrsim 20$ km from the SP (again see App. \ref{appendix1}). 
Computationally, our spectral multi-grid method performs quite well: Considering the above setup, the full solution
and derived quantities such as the stress tensor are calculated on a timescale of less than half an hour on an average machine
using a single processor.

\subsection{Numerical results}
\begin{figure}
\centering
\includegraphics[angle=0,width=1\linewidth]{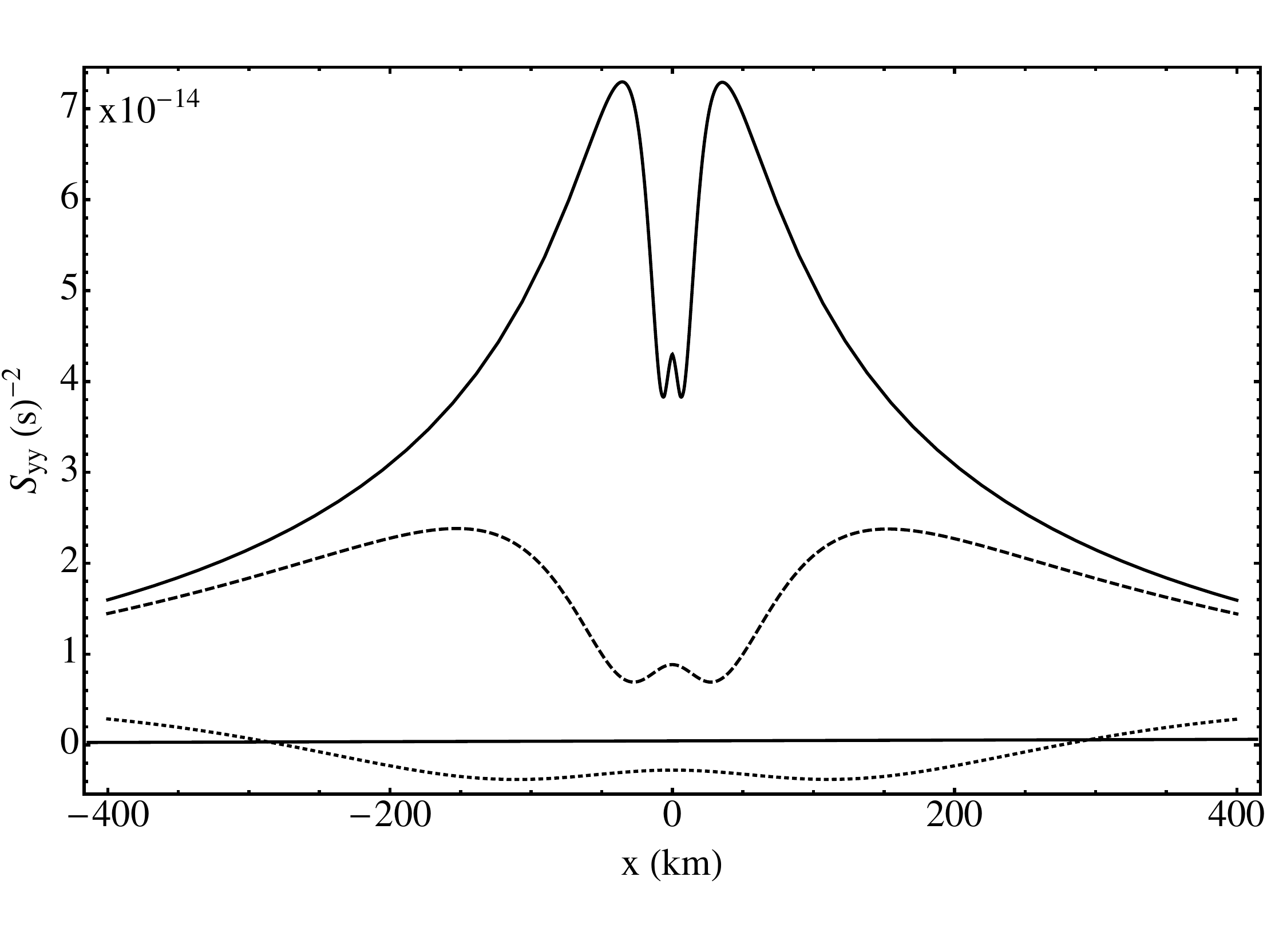}
\caption{Two-body QMOND tidal stress signal for $z=0$ along the lines $y\approx 25$ (solid line), $y\approx 100$ (dashed line) and $y\approx 400$ km (dotted line) inside the Earth-Sun MOND bubble: For the calculation, we have assumed the interpolation function Eq. \eqref{eq:m11}.
At $y\approx 25$ ($y\approx 400$) km, the positive (negative) peak is at $S_{yy}\approx 7.5\times 10^{-14}{\rm s}^{-2}$
($S_{yy}\approx -0.2\times 10^{-14}{\rm s}^{-2}$).}
\label{fig:14}
\end{figure}

\begin{figure}
\centering
\includegraphics[angle=0,width=1\linewidth]{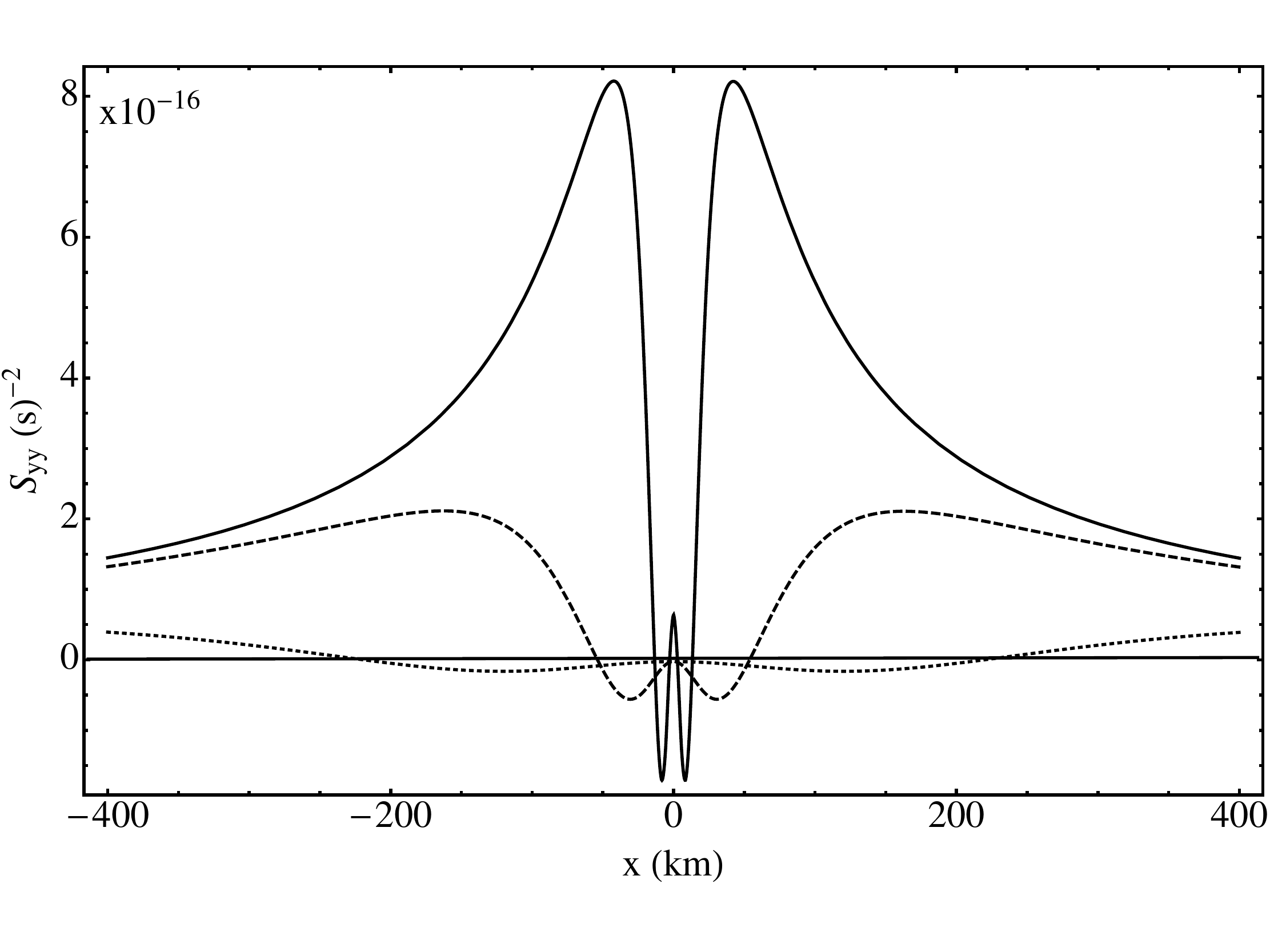}
\caption{Same as Fig. \ref{fig:14}, but now assuming the simple interpolation function Eq. \eqref{eq:m10}:
At $y\approx 25$ ($y\approx 400$) km, the positive (negative) peak is at $S_{yy}\approx 8\times 10^{-16}{\rm s}^{-2}$
($S_{yy}\approx -0.1\times 10^{-16}{\rm s}^{-2}$).}
\label{fig:16}
\end{figure}

\begin{figure}
\centering
\includegraphics[angle=0,width=1\linewidth]{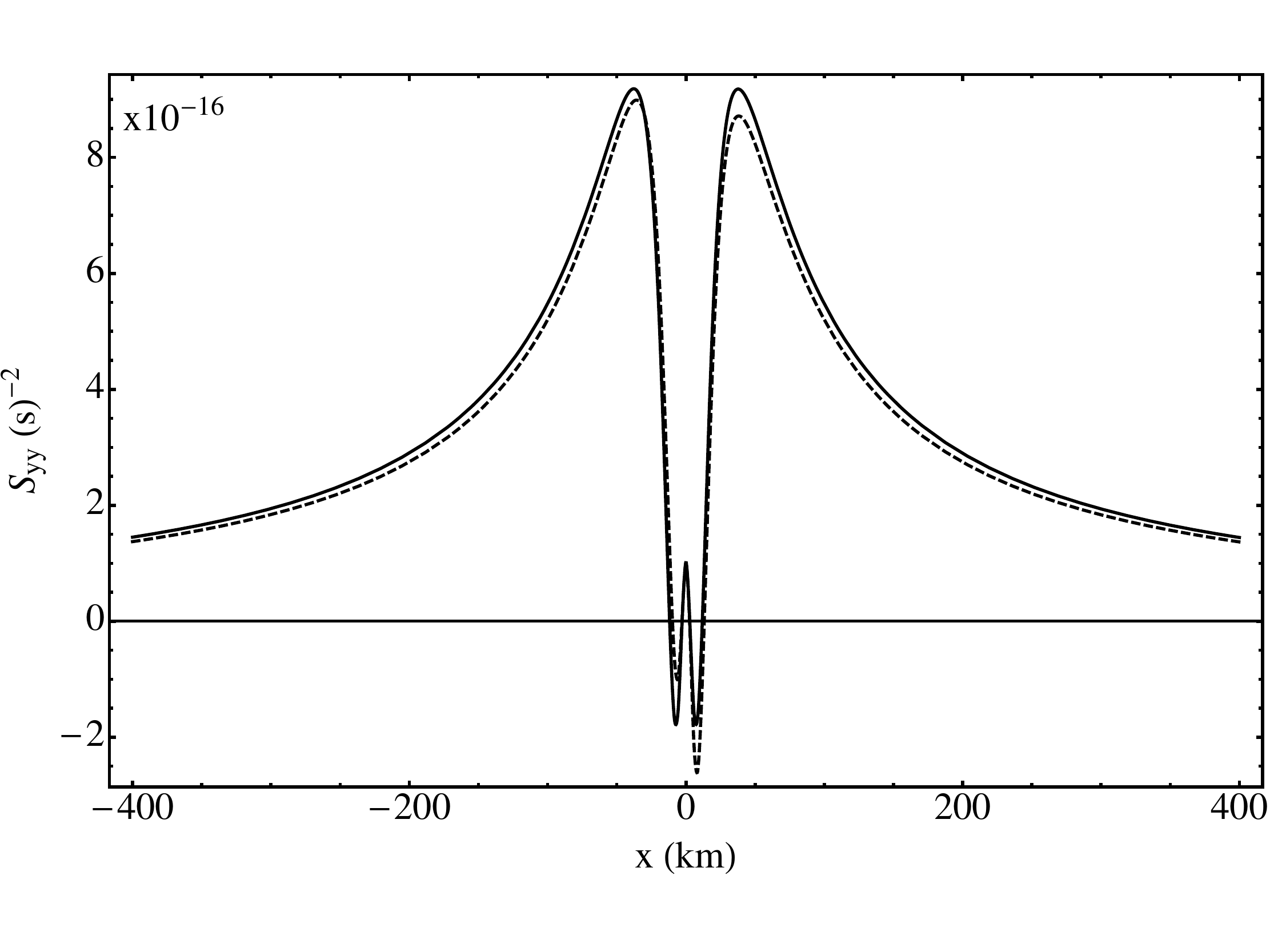}
\caption{Three-body QMOND tidal stress signal within the Earth-Sun bubble for $z=0$ along the line $y\approx 25$ km for different lunar phases, computed using the simple interpolation function Eq. \eqref{eq:m10}: The solid (dotted) line corresponds to the tidal stress signal at approximately full (new) moon. Note that the presence of the Moon has only a slight impact on the signal.}
\label{fig:17}
\end{figure}

Using the above multi-grid FFT code, we have numerically solved Eq.\eqref{eq:m1} near the Earth-Sun saddle point. The input PDM densities have been calculated for the two-body (Earth-Sun) and the three-body (Earth-Sun-Moon) approximations. Hereafter, we adopt a reference frame centered on the SP, with the $x$-axis pointing along the Earth-Sun direction (increasing toward the Sun). Note that within this reference frame, which is identical to the one used in Ref. \cite{bevis2010}, the LPF instrument will only be sensitive to the $S_{yy}$ component of the stress tensor because of certain design characteristics \cite{trenkel2010}. 
The results of our simulations for the Earth-Sun bubble are shown in Figs. \ref{fig:14}, \ref{fig:16} and \ref{fig:17}:  Using the Bekenstein-like interpolation function given by Eq. \eqref{eq:m11}, Fig. \ref{fig:14} shows the two-body QMOND tidal stress signal for $z=0$ at cuts of around $y=25$, $100$ and $400$ ($\pm 2.8$ due to the fixed sampling on equidistant grids) km away from the Earth-Sun SP. Our result fairly reproduces that of Ref. \cite{bevis2010} (cf. their Fig. 6). Compared to the calculation for TeVeS, the main differences are a generally larger tidal stress signal at all cuts (by about a factor of $1.7$) and a deeper central wiggle. Note that this agrees well with the conclusions of Ref. \cite{magueijo2011} about ``Type II'' MOND-like theories and the softening effect of the TeVeS curl field.

We have also computed the tidal stress signal using the simple interpolation function Eq. \eqref{eq:m10}. As expected from the earlier analysis of
the PDM density, the tidal stress signal decreases by approximately two orders of magnitude in this case. From Fig. \ref{fig:16}, one finds that the peak value of the tidal stress at $z=0$ and $y\approx 25$ km is approximately $8\times10^{-16}{\rm s}^{-2}$. Including the Moon into the calculation
(see Fig. \ref{fig:17}) has only a small impact on the estimated stress signal. We have also verified that the tidal stress signal becomes even lower ($S_{yy}\approx 10^{-19}{\rm s}^{-2}$ at $z=0$ and $y\approx 25$ km) if the ``fast-transition'' interpolation function Eq. \eqref{eq:m12} is used. In the next section, we shall discuss the predicted additional stress signal in the light of the expected
LPF sensitivity.

\subsection{LPF sensitivity and the QMOND signal} 
Choosing the analog of the Bekenstein-like $\nu$-function Eq. \eqref{eq:m11} in the framework of TeVeS, it was concluded that the instrument onboard the LPF spacecraft should have enough sensitivity to detect the additional MONDian tidal stress signal \cite{bekenstein2006,bevis2010,magueijo2011}. At a distance from the SP similar to the predicted impact parameter, this signal was estimated to be on the order of $S_{yy}\approx 10^{-14}{\rm s}^{-2}$. As we have seen in the previous sections,
however, the expected signal strength is very sensitive to the specific form of $\nu(y)$, and one might question whether such a conclusion remains valid for other choices. 
In particular, we will be interested in the situation for interpolation functions like Eq. \eqref{eq:m10} which provide a good description of the phenomenology on extragalactic scales.

Apart from the instrumental characteristics, the sensitivity of the LPF probe to tidal stresses depends on the frequency of the MOND signal
\footnote{See Ref. \cite{magueijo2011} for a detailed explanation of how the LPF sensitivity is estimated and a discussion about the impact of the spacecraft's flyby speed and self gravity on the tidal stress measurement.}. When the LPF spacecraft crosses the Earth-Sun bubble, the frequency scale of the MOND signal will be given by
$f = v/l_{b}$, where $v$ is the spacecraft's velocity and $l_{b}$ denotes the characteristic length scale of variations in the MOND tidal stress.
It can be seen from Figs. \ref{fig:14} and \ref{fig:16} that the QMOND tidal stress signal typically varies over a length of $l_{b}\approx 100$ km
for both choices of the interpolation function. Note that this scale length is the same as the one adopted in Ref. \cite{magueijo2011}. Given a
model for the power spectral density (PSD) $S_{h}(f)$ of the noise, the expected signal-to-noise ratio ($S/N$) may be roughly estimated from the
signal's characteristic amplitude $h_{c}$, $S/N\sim h_{c}/\sqrt{fS_{h}}$ \cite{gwaves2009}. To obtain a sensitivity threshold for the LPF, we set
$S/N\sim 1$. Considering $v=1.5$ km s$^{-1}$ \footnote{Note that typical LPF passing velocities are expected to fall within the range $1<v<3$
km s$^{-2}$, which is already near optimal regarding the expected $S/N$ ratios \cite{magueijo2011}.} and a constant noise PSD with
$\sqrt{S_{h}}\approx 1.5\times 10^{-14}{\rm s}^{-2}/\sqrt{\rm Hz}$ \cite{magueijo2011}, this gives $f\approx 10^{-2}$ Hz and yields a
threshold on the order of $10^{-14}{\rm s}^{-2}$.
This should be sufficient to detect the QMOND tidal stress signal for interpolation functions like Eq. \eqref{eq:m11}
whose amplitude lies between $2\times 10^{-14}$ and $8\times 10^{-14}$ s$^{-2}$ at $y=50$ km (see Fig. \ref{fig:14}). For the simple function
Eq. \eqref{eq:m10}, however, the situation is different: Since the tidal stress signal decreases by approximately two orders of magnitude, we
expect that the signal will not be detectable at the given impact parameter. Therefore, to have a chance of probing these models, the LPF
spacecraft might need to pass the Earth-Sun SP much closer than previously thought, but it is also possible that the modified signal remains
completely undetectable.

To get more insight on the above, we follow the lines of Ref. \cite{magueijo2011} and consider the signal's amplitude spectral density (ASD)
which is given by the square root of the PSD:
\beq
P(f) = \frac{2}{T}\left\lvert\int_{-T/2}^{+T/2}h(t)\exp(-2\pi ift)dt\right\rvert^{2},
\label{eq:m13}
\eeq
where $f$ is the frequency, $h(t)=S_{yy}(vt,b,0)$ corresponds to the signal, and $b$ is the impact parameter, i.e. the minimum distance from
the SP which is approached at $t=0$. Fixing $v=1.5$ km s$^{-1}$ \cite{bevis2010,magueijo2011}, we furthermore have the integration period $T$
which we choose as $T\approx 1.3\times 10^{4}$s, in accordance with our numerical setup (see Sec. \ref{section4}). To compare the signal ASD
to that of the noise, we use the simplified LPF noise model discussed in Ref. \cite{trenkel2010} which assumes a constant baseline within
$1<f<10$ mHz, with an ADS around $1.5\times 10^{-14}{\rm s}^{-2}/\sqrt{\rm Hz}$. Beyond this range, we assume that the sensitivity degrades
with $1/f$ and $f^{2}$ for lower and higher frequencies, respectively. Note that although this simple noise model will suffice for obtaining
realistic estimates, improved LPF noise models have recently been presented in Ref. \cite{magueijo2011} (see their Fig. 6).

\begin{figure}
\centering
\includegraphics[angle=0,width=1\linewidth]{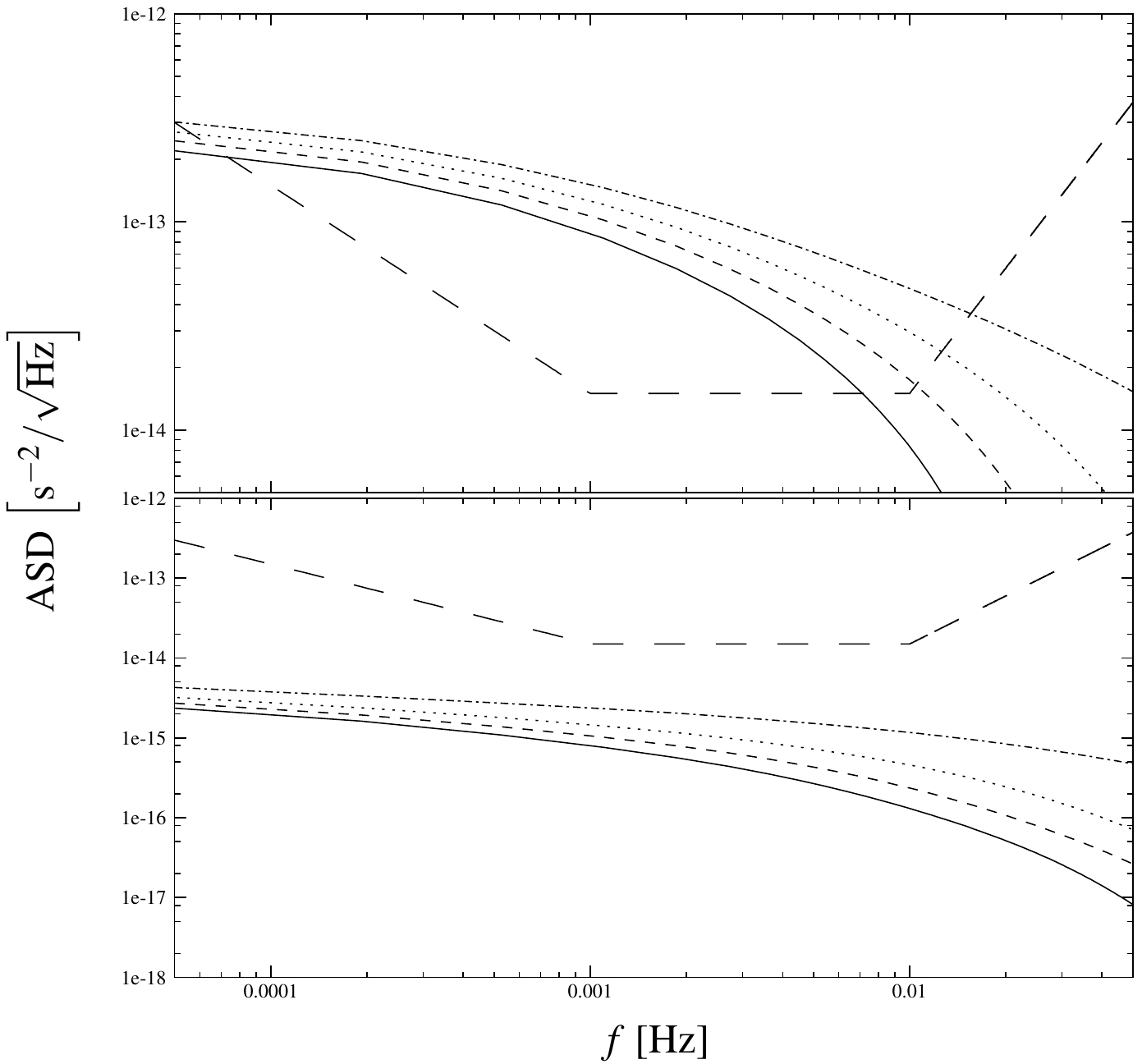}
\caption{{\it Top panel:} The ASD of the MONDian tidal stress signal for $\nu(y)$ given by Eq. \eqref{eq:m11} compared to the simple noise model
described in the text; assuming $v=1.5$ km s$^{-1}$, we illustrate results for $b\approx 45$ (solid line), $25$ (dashed line), $12$ (dotted line)
and $1$ km (dot-dashed line). {\it Bottom panel:} Same as the top panel, but now assuming the simple interpolation function Eq. \eqref{eq:m10}.}
\label{fig:sn}
\end{figure}

For the interpolation function Eq. \eqref{eq:m11}, the top panel of Fig. \ref{fig:sn} illustrates the resulting ASDs of the signal, assuming different
impact parameters between approximately $45$ and $1$ km. Since the numerically calculated stress becomes unreliable for distances $\lesssim 20$ km
from the SP (see App. \ref{appendix1}), our results for small impact parameters underestimate the signal at the high-frequency end, i.e. for
$f\gtrsim 7\times 10^{-2}$ Hz. However, this corresponds to a frequency domain where the noise is dominant, and thus we expect this resolution
effect to have no significant impact on our conclusions. As can be seen from the figure, $S/N$ is typically on the order of $10$ per $\sqrt{Hz}$ over a broad
frequency range. Applying the techniques of noise matched filtering (which are frequently used in the search for gravitational waves) to the present
problem, it can further be shown \cite{gwaves2009,magueijo2011} that the maximal $S/N$, realized by an optimal filtering template, is given by
\beq
(S/N)_{\rm opt} = 2\left\lbrack\int_{0}^{\infty}df\frac{\left\lvert\tilde{h}(f)\right\rvert^{2}}{S_{h}(f)}\right\rbrack^{1/2}.
\label{eq:m13b}
\eeq
Here $S_{h}(f)$ again denotes the PSD of the noise and $\tilde{h}(f)$ is the Fourier transform of the signal. Using Eq. \eqref{eq:m13b}, the
above scenario for the Bekenstein-like interpolation function Eq. \eqref{eq:m11} translates into integrated $S/N$ ratios of around $50-105$,
which is in good agreement with the TeVeS results of Ref. \cite{magueijo2011}.

In the bottom panel of Fig. \ref{fig:sn}, we show the corresponding signal ASDs for the simple interpolation function Eq. \eqref{eq:m10}.
Almost throughout the entire frequency range, the calculated signals now appear by at least an order of magnitude below the noise level.
For an impact parameter of $b\approx 12$ km and larger, this gives rise to an integrated $S/N$ of around $1$ or less. At $b\approx 45$ km, for
instance, we have $S/N\approx 0.5$. Only in the case of $b\approx 1$ km, we find that the maximal $S/N$ crosses unity, taking a value of $2$.
To test models based on interpolation functions similar to Eq. \eqref{eq:m10}, the LPF spacecraft would therefore be required to pass the
Earth-Sun SP at a distance on the order of a $1$ km or less. Note, however, that since the expected integrated $S/N$ ratios are rather close
to unity, these estimates are quite sensitive to the details of the true underlying noise model which is currently unknown. Depending on the
actual noise model, this could, in principal, give rise to exposing the modified signal to a wider radial range or making it virtually
undetectable. Also, considering potentially necessary impact parameters $b\lesssim 1$ km, practical issues related to positional accuracies
as well as the spacecraft's self-gravity might be of concern \cite{magueijo2011}. Taking all of this into account, it seems safe to conclude
that the simple interpolation function Eq. \eqref{eq:m10} has a good chance of surviving a LPF null result. If the ``fast-transition'' interpolation
function Eq. \eqref{eq:m12} is adopted, then the $S/N$ ratios are even further suppressed, and the modified tidal stress signal will basically
become invisible to the LPF. Finally, note that we have also considered the framework of TeVeS and addressed the question whether there exist viable interpolation functions which give rise to similar conclusions. For this purpose, we have used our previous choices of $\nu(y)$ to formally construct the
corresponding functions $\hat{\mu}$ for the TeVeS scalar field. A detailed description of this procedure and the implications of our results are
separately discussed in App. \ref{appendix2}.

\section{Conclusions}
\label{section5} 
Low-acceleration regions encasing the SPs of the gravitational potential in the SS, may exhibit significant deviations from Newton's laws within
the framework of MOND or closely related theories. As the LPF spacecraft may visit the Earth-Sun low-acceleration bubble in the near future, this
could provide a unique occasion to put MOND-like modifications of gravity to a direct test. While it is known that the quoted LPF sensitivity is
good enough to detect the non-Newtonian tidal stress signal in certain cases, it is still very important to explore the full space of theoretical
models and possible outcomes of the planned flyby experiment. Also, a necessary prerequisite for any realistic performance of this and 
future experiments, is the precise knowledge of the positions and motions of such bubbles near the Earth.

Using a high accuracy model of the SS Newtonian potential, we have determined the positions and tracked the motions of the Earth-Sun, the Moon-Sun
and Saturn-Sun SPs with an accuracy of $\Delta_{\rm E-S}\approx 4$ m and $\Delta_{\rm Sat-S}\approx 2$ km for the Earth-Sun SP and Saturn's SP respectively.
During the analysis of the SP's motion we have further found that a subset of Saturn's satellites, consisting of Surtur, Ymir, Fenrir and Loge, all regularly pass through the Saturn-Sun MOND bubble. This provides the first example of SS objects periodically entering the MOND regime. An accurate study of the motions of these satellites, may in principle be used as an independent test of the MOND paradigms, but clearly more investigations are needed. We plan to perform a quantitative analysis on how the MONDian perturbations impact the satellites' trajectories in a forthcoming paper.

Adopting the framework of QMOND, we have calculated the additional tidal stress signal due to the PDM density near the Earth-Sun SP. Using a Bekenstein-like interpolation function and assuming an impact parameter $b\approx 50$ km, we obtain an approximately by a factor of $1.7$ higher
tidal stress signal than estimated in the context of TeVeS, which is in accordance with previous considerations \cite{magueijo2011}. Choosing the QMOND analog of the
so-called simple interpolation function, which provides a good description of extragalactic phenomenology, we encounter a different situation.
In this case, the signal turns out to be significantly smaller and, taking the sensitivity of the LPF probe into account, we typically obtain
maximal (integrated) $S/N$ ratios on the order of $1$ or less. Although moving to very small impact parameters on the order of $\lesssim 1$ km
might, in principle, help to detect the anomalous stress signal, it seems safe to conclude that interpolation functions similar to the simple one
have a good chance of surviving a LPF null result. For interpolation functions with a ``faster'' transition from Newtonian dynamics to MONDian
behavior, the $S/N$ ratios are even further suppressed, and the modified tidal stress signal will basically become invisible to the LPF.

Finally, we want to point out that our conclusions do not necessarily apply to other proposed MOND frameworks. Considering the original form of TeVeS, for
instance, a transition corresponding to Eq. \eqref{eq:m12} yields an unacceptable interpolation function $\hat{\mu}$ for the scalar field, which is briefly discussed
in App. \ref{appendix2}. Moreover, further constraints on the rapidity of $\nu(y)$ in QMOND might arise once the full cosmological scenario of its associated relativistic
bimetric theory has been worked out.

\begin{acknowledgments}
P.G. wants to thank Marco Friuli for useful discussions. The authors want to thank the anonymous referee, Joao Magueijo and Mordehai Milgrom for useful discussions. M.F. is supported in part at the Technion by a fellowship from the Lady Davis Foundation. 
\end{acknowledgments}

\appendix

\section{Semi-analytic solutions near the SP}
\label{appendix1}
In the following, we shall consider the configuration of two bodies at distance $d$ with masses $M$ and $m$, respectively,
where we assume that the line connecting the bodies coincides with the $z$-axis. Shifting the coordinate origin to the SP
and introducing spherical polar coordinates with $z=r\cos\theta$, the Newtonian potential in the vicinity of the saddle
approximately takes the form
\begin{equation}
\Phi_{N} = Cr^{2}\left (1-3\cos^{2}\theta\right ) + {\rm const},
\label{eq:a1}
\end{equation}
where
\begin{equation}
\begin{split}
C = \frac{G}{2d^{3}} & \left\lbrack M\left (\frac{M-\sqrt{Mm}}{M-m}\right )^{-3}\right. \\
{ } & + m \left.\left ( 1 - \frac{M-\sqrt{Mm}}{M-m}\right )^{-3}\right\rbrack .
\end{split}
\label{eq:a2}
\end{equation}
If $M\gg m$ (as in case of the Earth-Sun SP, for instance), Eq. \eqref{eq:a2} may be well approximated by
\begin{equation}
C \approx \frac{GM}{2d^{3}}\left (4+\sqrt{\frac{M}{m}}\right ).
\label{eq:a2b}
\end{equation}
Near the SP where $\Delta\Phi_{N}=0$, the QMOND contribution to the total gravitational potential $\Phi$,
\begin{equation}
\Phi = \Phi_{N} + \Phi_{QM},
\label{eq:a3}
\end{equation}
is associated with the phantom source distribution
\begin{equation}
\tilde{\rho}_{QM} \equiv \Delta\Phi_{QM} = \bm{\nabla}\cdot\left\lbrack\nu(y)\bm{\nabla}\Phi_{N}\right\rbrack.
\label{eq:a4}
\end{equation}
Although Eq. \eqref{eq:a4} does generally not exhibit analytic solutions, it is possible to find approximate expressions
for certain cases, which we shall discuss in the following sections.

\subsection{Quasi-Newtonian domain}
If the resulting dynamics is sufficiently close to the Newtonian limit, i.e. $y=g_{N}/a_{0}\gg 1$, we may expand the interpolating function $\nu(y)$ in powers of $y$,
\begin{equation}
\nu(y) \approx 1 + \frac{\alpha_{1}}{y} + \frac{\alpha_{2}}{y^{2}} + \dots .
\label{eq:a5}
\end{equation}
Substituting the above into Eq. \eqref{eq:a4} and remembering that $\Delta\Phi_{N}=0$ near the saddle, we thus obtain
\begin{equation}
\begin{split}
\Delta\Phi_{QM} &\approx \alpha_{1}\bm{\nabla}\cdot\left (\frac{\bm{\nabla}\Phi_{N}}{y}\right ) + \alpha_{2}\bm{\nabla}\cdot\left (\frac{\bm{\nabla}\Phi_{N}}{y^{2}}\right ) + \dots\\
&\equiv \tilde{\rho}^{(1)}_{QM} + \tilde{\rho}^{(2)}_{QM} + \dots .
\end{split}
\label{eq:a6}
\end{equation}
The linearity of Eq. \eqref{eq:a6} allows one to solve the equation order by order. Using Eq. \eqref{eq:a1}, one finds that the $n$th-order
contribution to the source term can be written as
\begin{equation}
\tilde{\rho}^{(n)}_{QM} = -\frac{A_{n}}{r^{n}}\frac{1-9\cos^{2}\theta}{\left (1+3\cos^{2}\theta\right )^{(n+2)/2}},\quad n\in\mathbb{N},
\label{eq:a7}
\end{equation}
where we have defined
\begin{equation}
A_{n} = \frac{n\alpha_{n}a_{0}^{n}}{\left (2C\right )^{n-1}}.
\label{eq:a7a}
\end{equation}
Let us begin with the first-order contribution.
As we are only interested in the saddle region, the approximate Newtonian potential in Eq. \eqref{eq:a1} suggests an ansatz of the form
\begin{equation}
\Phi_{QM}^{(1)}(r,\theta ) = \gamma_{1}r^{\beta_{1}}\mathcal{T}^{(1)}(\theta) + {\rm const},
\label{eq:a8}
\end{equation}
where $\beta_{1}$ and $\gamma_{1}$ are assumed as constant.
Inserting the above expression into Eq. \eqref{eq:a6} then fixes $\beta_{1} = 1$ and $\gamma_{1} = A_{1}$, leaving us with a single
ordinary differential equation for $\mathcal{T}^{(1)}(\theta)$,
\begin{equation}
\begin{split}
2&\mathcal{T}^{(1)}(\theta) + \frac{\cos\theta}{\sin\theta}\frac{d}{d\theta}\mathcal{T}^{(1)}(\theta) + \frac{d^{2}}{d\theta^{2}}
\mathcal{T}^{(1)}(\theta) \\
{ } &= -\frac{1-9\cos^{2}\theta}{\left (1+3\cos^{2}\theta\right )^{3/2}}.
\end{split}
\label{eq:a9}
\end{equation}
For a physically relevant solution, we require that $\mathcal{T}^{(1)}(\theta)$ is smooth and periodic on the interval $\lbrack 0,\pi\rbrack$ (see, e.g., \cite{bekenstein2006}). In this case, Eq. \eqref{eq:a9} may numerically be solved in terms of a Fourier series, yielding
\begin{equation}
\begin{split}
\mathcal{T}^{(1)}(\theta) &\approx 0.06310 - 0.23719\cos 2\theta\\
{ } &+ 0.02659\cos 4\theta - 0.00498\cos 6\theta + \dots .
\end{split}
\label{eq:a10}
\end{equation}
The above series converges rather quickly and may already be cut off after the first eight terms to achieve a relative
accuracy of $\lesssim 10^{-4}$.

Next, we shall consider the second-order correction to the potential $\Phi_{QM}$. It turns out that one cannot adopt the same ansatz as in
Eq. \eqref{eq:a8} because it is impossible to meet the smoothness condition of the angular part in this case. As we will see shortly, this can
easily be remedied by considering an additional term proportional to $\log (r/r_{s})$ in the potential. Thus we start from
\begin{equation}
\begin{split}
\Phi_{QM}^{(2)}(r,\theta ) = \gamma_{2}\left\lbrack r^{\beta_{2}}\mathcal{T}^{(2)}(\theta) - B\log\left (\frac{r}{r_{s}}\right )\right\rbrack + {\rm const}.
\end{split}
\label{eq:a11}
\end{equation}
Again substituting this expression into Eq. \eqref{eq:a6}, we find that $\beta_{2} = 0$ and $\gamma_{2} = A_{2}$. This requires the constant $B$ to be dimensionless and gives rise to the following equation for $\mathcal{T}^{(2)}(\theta)$:
\begin{equation}
\begin{split}
\frac{\cos\theta}{\sin\theta}\frac{d}{d\theta}\mathcal{T}^{(2)}(\theta) + \frac{d^{2}}{d\theta^{2}}\mathcal{T}^{(2)}(\theta) =  -\frac{1-9\cos^{2}\theta}{\left (1+3\cos^{2}\theta\right )^{2}} + B.
\end{split}
\label{eq:a12}
\end{equation}
Since only derivatives of $\mathcal{T}^{(2)}(\theta)$ appear in Eq. \eqref{eq:a12}, one may integrate once and finally obtain
\begin{equation}
\begin{split}
\frac{d}{d\theta}\mathcal{T}^{(2)}(\theta) = \frac{1}{\sin\theta}&\left\lbrack\frac{2\cos\theta}{1+3\cos^{2}\theta} - B\cos\theta + D\right.\\
{ } - &\left. \frac{\sqrt{3}}{3}\arctan\left (\sqrt{3}\cos\theta\right )\right\rbrack , 
\end{split}
\label{eq:a13}
\end{equation}
where $D$ is an integration constant. Applying the periodicity and smoothness conditions to the above, a brief calculation leads to $D=0$ and
\begin{equation}
B = \frac{1}{2} - \frac{\sqrt{3}}{9}\pi .
\label{eq:a14}
\end{equation}
Once again, an approximate solution of $\mathcal{T}^{(2)}(\theta)$ may be found by using a Fourier series approach, and the resulting series,
\begin{equation}
\begin{split}
\mathcal{T}^{(2)}(\theta) &\approx - 0.10460\cos 2\theta + 0.02177\cos 4\theta\\
{ } &- 0.00524\cos 6\theta + 0.00136\cos 8\theta + \dots ,
\end{split}
\label{eq:a15}
\end{equation}
quickly converges toward the real solution. In principle, one could apply the general procedure outlined in this section also to higher orders,
but for the purpose of this paper, we shall not take these into account. Finally, note that the semi-analytic solutions found in this section automatically satisfy the desired Newtonian boundary conditions because of $\beta_{1},\beta_{2} < 2$, meaning that the total gravitational
potential is entirely dominated by the expression in Eq. \eqref{eq:a1} when moving to radii much larger than the true size of the bubble,
i.e. $r/r_{0}\gg 1$, where $r_{0}=r_{\rm trig}$ for the interpolation function defined in Eq. \eqref{eq:m11}. 

\subsection{Deep-MOND domain}
In the limit of very small gravitational fields, we have $y\rightarrow 0$ and the QMOND interpolating function may be approximated as
\begin{equation}
\nu(y) \approx \frac{1}{\sqrt{y}},
\label{eq:a16}
\end{equation}
which is independent of the particularly chosen form of $\nu(y)$.
Using Eq. \eqref{eq:a16} in Eq. \eqref{eq:a4}, the corresponding phantom source term can be expressed as
\begin{equation}
\tilde{\rho}_{QM} = -\frac{A}{\sqrt{r}}\frac{1-9\cos^{2}\theta}{\left (1+3\cos^{2}\theta\right )^{5/4}},
\label{eq:a17}
\end{equation}
where
\begin{equation}
A = \frac{1}{2}\sqrt{2Ca_{0}}.
\label{eq:a18}
\end{equation}
Similar to our approach in the quasi-Newtonian situation, we make the ansatz
\begin{equation}
\Phi_{QM}(r,\theta ) = \gamma r^{\beta}\mathcal{T}(\theta ) + {\rm const}
\label{eq:a19}
\end{equation}
which is again substituted into Eq. \eqref{eq:a4}. This time we obtain $\beta = 3/2$ and $\gamma = A$, with the resulting differential equation for
$\mathcal{T}(\theta )$ given by
\begin{equation}
\begin{split}
\frac{15}{4}&\mathcal{T}(\theta) + \frac{\cos\theta}{\sin\theta}\frac{d}{d\theta}\mathcal{T}(\theta) + \frac{d^{2}}{d\theta^{2}}\mathcal{T}(\theta)\\
{ } &= -\frac{1-9\cos^{2}\theta}{\left (1+3\cos^{2}\theta\right )^{5/4}}.
\end{split}
\label{eq:a20}
\end{equation}
Analogously to the first-order contribution in the quasi-Newtonian domain, an approximate solution is determined by expanding $\mathcal{T}(\theta )$ into a Fourier series on the interval $\lbrack 0,\pi\rbrack$. A straightforward calculation then yields
\begin{equation}
\begin{split}
\mathcal{T}(\theta) &\approx -0.06750 - 0.53716\cos 2\theta\\
{ } &+ 0.03017\cos 4\theta - 0.00476\cos 6\theta + \dots ,
\end{split}
\label{eq:a21}
\end{equation}
where it is sufficient to consider the first six terms of Eq. \eqref{eq:a21} for an accuracy of $\lesssim 10^{-4}$. It is important to
note that Eq. \eqref{eq:a19} merely constitutes a special solution to Eq. \eqref{eq:a4}, and thus we still need to address the question
of boundary conditions in the deep-MOND case.

Given a suitable choice of such boundary conditions, Eq. \eqref{eq:a19} needs to be replaced by 
\begin{equation}
\Phi_{QM}(r,\theta ) = Ar^{3/2}\mathcal{T}(\theta ) + \Phi_{0}(r,\theta ) + {\rm const},
\label{eq:a21b}
\end{equation}
where $\Phi_{0}(r,\theta )$ is an appropriate solution of Laplace's equation. Using the previously assumed smoothness and periodicity
condition of the angular part and additionally requiring that the potential is regular at $r=0$, the most general form of $\Phi_{0}(r,\theta )$
can be expressed as
\begin{equation}
\Phi_{0}(r,\theta ) = \sum\limits_{n=1}^{\infty} q_{n}\frac{C^{2n-1}}{a_{0}^{2n-2}}r^{2n}P_{2n}(\cos\theta ),\quad n\in\mathbb{N},
\label{eq:a21c}
\end{equation}
where $C$ is defined according to Eq. \eqref{eq:a2}, $P_{n}$ denotes the Legendre polynomial of degree $n$ and $q_{n}\in\mathbb{R}$.
Since the validity of equation \eqref{eq:a21b} is restricted to
a close vicinity of the SP at $r=0$, however, we expect that only the first few terms in Eq. \eqref{eq:a21c} will significantly contribute
to the solution in this case. As we shall see later, this allows one to approximate the infinite series in Eq. \eqref{eq:a21c} by a cut-off one. 
Unfortunately, there exists no independent way of determining the coefficients $q_{n}$ because the saddle region is devoid of any real sources.
Thus their values are set by the boundary conditions in the intermediate MOND domain and need to be estimated from the full numerical result,
which is further discussed in the next section.

\begin{figure}
\centering
\includegraphics[width=1\linewidth ]{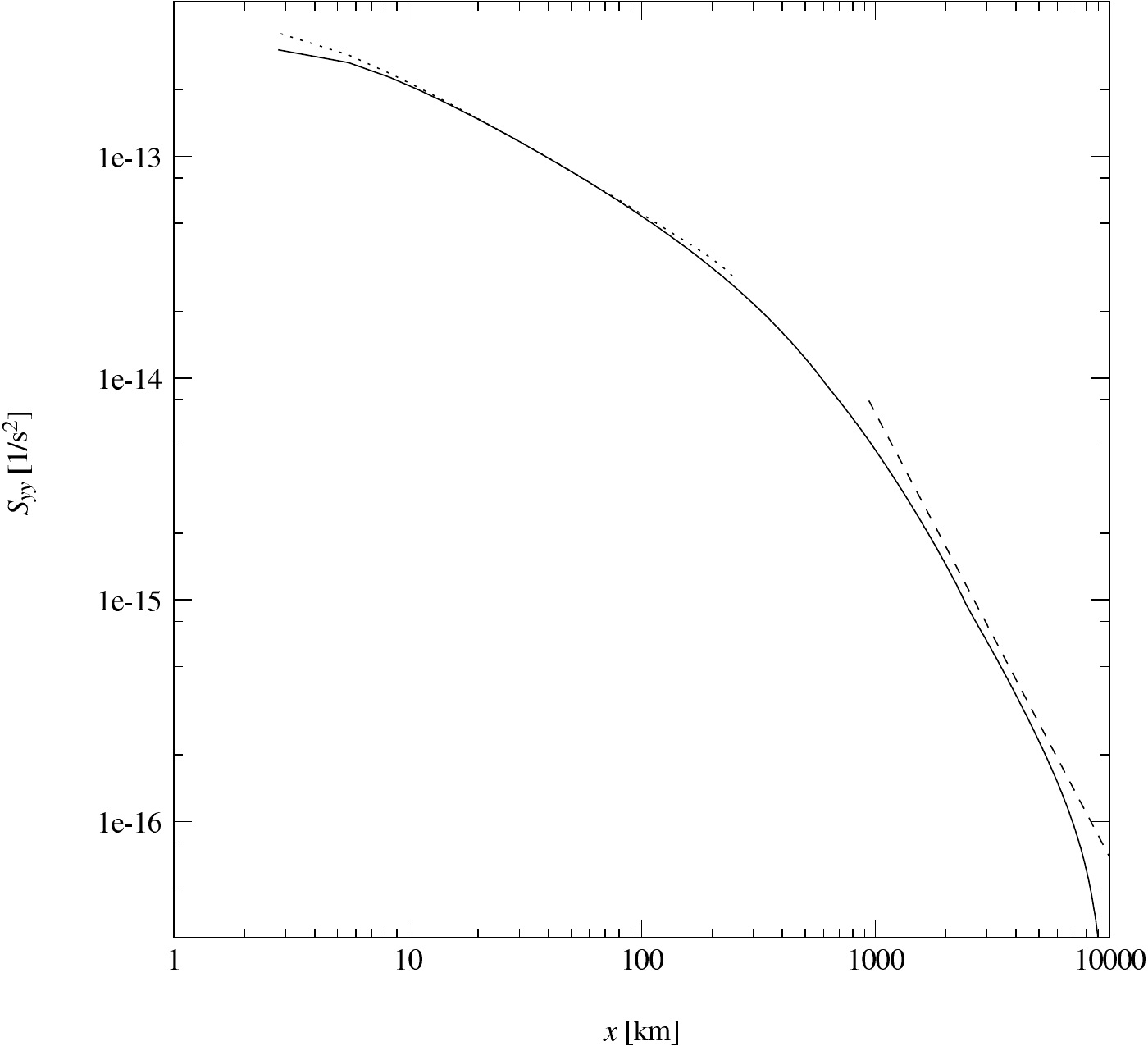}
\caption{Potential derivative $S_{yy}$ of the Earth-Sun SP evaluated at $y\approx 1.4$ and $z=0$ in units of km: Shown are the numerical result
for $\nu(y)$ given by Eq. \eqref{eq:m11} with $k=0.03$ (solid line) as well as the semi-analytic approximations in the quasi-Newtonian (dashed line)
and deep-MOND (dotted line) regime, respectively.}
\centering
\label{fig:a2}
\end{figure}

\subsection{Comparison to numerical results}
Having derived semi-analytic solutions in both the quasi-Newtonian and the deep-MOND domain, we now want to compare these to the
full numerical two-body results of the Earth-Sun SP. To do so, however, we still need to determine the series coefficients in Eq. \eqref{eq:a5} for
a particular choice of $\nu(y)$. Let us consider the expression in Eq. \eqref{eq:m11} which is the QMOND analog of the TeVeS interpolating function
used in Ref. \cite{bevis2010}. For $y\gg 1$, the above formula can be expanded as
\begin{equation}
\nu(y) = 1 + \frac{1}{4}\left (\frac{4\pi}{k}\right )^{3}\frac{1}{y^{2}} + \mathcal{O}\left (\frac{1}{y^{4}}\right ),
\label{eq:a23}
\end{equation}
which reveals that the first-order term vanishes, i.e. one has $\alpha_{1} = 0$.
Since the numerical approach uses different boundary conditions (see Sec. IV B), we choose not to compare the actual potentials, but
rather their second derivatives. In accordance with our numerical setup, we switch to Cartesian coordinates, letting the symmetry axis now point
along the $x$-direction.

\begin{figure}
\centering
\includegraphics[width=1\linewidth ]{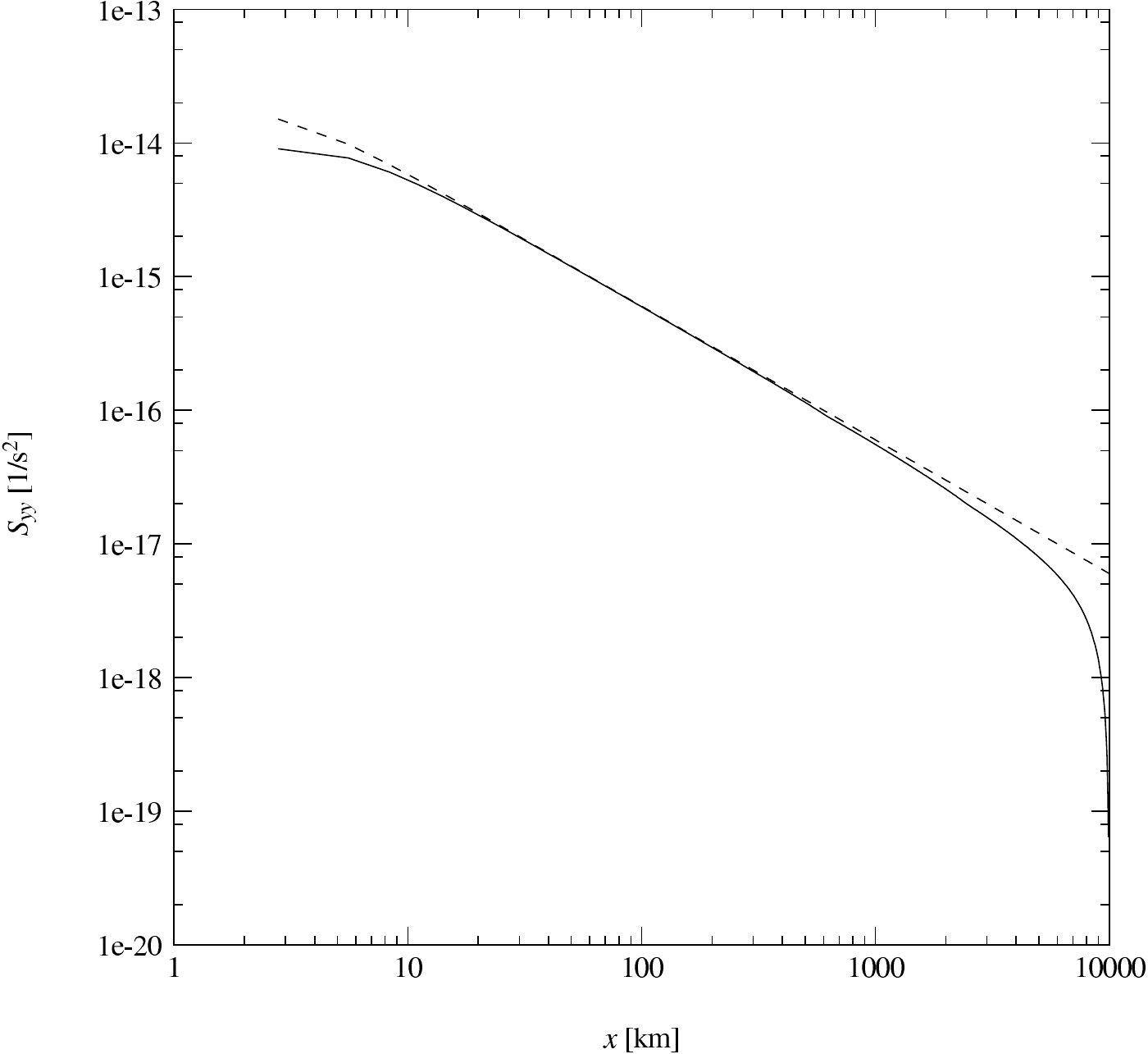}
\caption{Same as Fig. \ref{fig:a2}, but now for the function $\nu(y)$ given by Eq. \eqref{eq:m10}.}
\centering
\label{fig:a3}
\end{figure}

As for a comparison to the deep-MOND solution, we must also take care of the homogeneous contribution $\Phi_{0}$ which is determined by the
potential values in the intermediate MOND domain. The coefficients $q_{n}$ can be calculated by fitting the simulation results with the help
of Eq. \eqref{eq:a21b}, which, of course, requires introducing a cut-off in Eq. \eqref{eq:a21c}. Since only the first few terms significantly
contribute in the close vicinity of the SP, we cut the series off at $n=3$ and discard all remaining terms. Furthermore, to ensure that Eq.
\eqref{eq:a16} is approximately satisfied, we consider only data points with $0.05 < r/r_{0} < 0.5$ for this procedure \cite{bevis2010}.
Setting $k=0.03$ in Eq. \eqref{eq:m11} and using the corresponding simulation data of the Earth-Sun SP, we find
\begin{equation}
\begin{split}
q_{1} &\approx 2.001\times 10^{-3},\\
q_{2} &\approx -2.433\times 10^{-13}, \\
q_{3} &\approx 1.228\times 10^{-22}.
\end{split}
\label{eq:a23c}
\end{equation}
Fixing $y\approx 1.4$ and $z=0$ in units of km, the numerical two-body result for $S_{yy}\equiv \partial_{y}^{2}\Phi_{QM}$
of the Earth-Sun SP is illustrated in Fig. \ref{fig:a2}. Since the numerical result is essentially symmetric
with respect to $x=0$, we only consider the case $x>0$. As can be seen from the figures, the approximate analytic solutions are well recovered
by the full numerical result. Note that the increasing deviation from the quasi-Newtonian solution near the boundary at $x=10^{4}$km is entirely
due to the different choice of boundary conditions (see Sec. \ref{section4}). The numerical solution becomes unreliable for $x\lesssim 20$km, which
is a consequence of the grid's finite resolution.

\begin{figure}
\centering
\includegraphics[angle=0,width=1\linewidth]{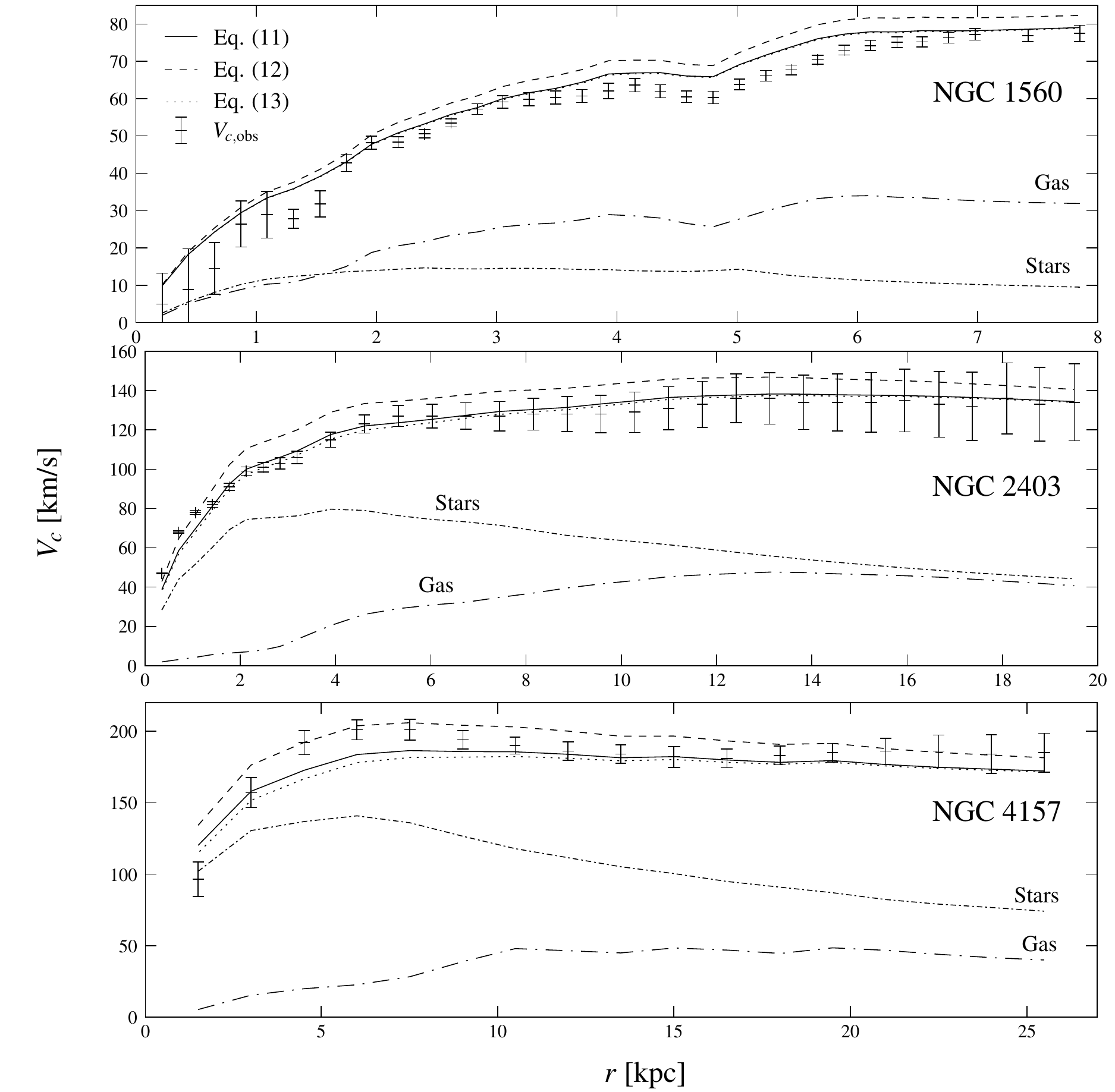}
\caption{Rotation curves of NGC 1560 (upper panel), NGC2403 (middle panel) and NGC 4157 (lower panel): hile The data points correspond to observations. The dashed,
solid and dotted lines represent rough ``fits'' obtained using Eq. \eqref{eq:m11} , Eq. \eqref{eq:m10} and Eq. \eqref{eq:m12} and setting $M/L=1$ (for NGC 1560 and
NGC 2403) and $M/L=1.5$ (for NGC 4157), respectively. Also shown are the Newtonian rotation curves of the individual stellar (dot-dashed line) and gas
(long dot-dashed line) components.}
\label{fig:rotcomp}
\end{figure}

Another example is given by the simple interpolating function defined in Eq. \eqref{eq:m10}.
Its expansion takes the form 
\begin{equation}
\nu(y) = 1 + \frac{1}{y} - \frac{1}{y^{2}} + \mathcal{O}\left (\frac{1}{y^{3}}\right ).
\label{eq:a25}
\end{equation}
From the above, one immediately identifies $\alpha_{1} = 1$ and $\alpha_{2} = -1$. As is shown in Fig. \ref{fig:a3} for $S_{yy}$, the quasi-Newtonian
solution provides an excellent approximation to the full numerical result in this case. On the other hand, the deep-MOND limit is not reached at the given resolution ($\Delta x_{\rm max}\approx 2.8$km).

\section{Impact of $\bm{\nu(y)}$ on galactic scales}
\label{appendixrot}
Any functional form of $\nu(y)$ chosen for the SS is required to be consistent with the vast amount of available extragalactic data. The
properties of Eqs. \eqref{eq:m10} and \eqref{eq:m11} relevant to these scales are well-known and have been extensively discussed in the literature
(see, e.g., Ref. \cite{zhao2006}). While both prescriptions are able to describe the observed rotation curves of spiral galaxies, it is usually concluded
that the data prefers the simple choice Eq. \eqref{eq:m10} (or further refined versions thereof) since it provides a better fit to observations if one leaves
the stellar mass-to-light ratio $M/L$ as a free parameter. For Eq. \eqref{eq:m12}, however, the situation is unclear. To get an idea about its performance, we
consider the rotation curves of three test galaxies covering different gravitational regimes: NGC 1560 (low surface brightness), NGC 2403 and NGC 4157 (high
surface brightness).
For the different choices of $\nu(y)$, Fig. \ref{fig:rotcomp} illustrates the predicted MOND rotation curves together with the observed ones, assuming constant
$M/L$ ratios in the B-band. Note that since we are only interested in the performance of Eq. \eqref{eq:m12} compared to the others, we have not tried to actually
fit the observed curves. Instead, to obtain a very rough match to the data, we have used $M/L=1$ for both NGC 1560 and NGC 2403 and $M/L=1.5$ for NGC 4157. As can
be seen from the figure, the behavior of Eq. \eqref{eq:m12} is very similar to that of the simple $\nu$-function, despite their differences on
SS scales. This suggests that the interpolating function Eq. \eqref{eq:m12} is not only compatible with extragalactic data, but also recovers the successful descriptive
power of Eq. \eqref{eq:m10}.

\section{Interpolating functions for TeVeS}
\label{appendix2}

\begin{figure}
\centering
\includegraphics[angle=0,width=1\linewidth]{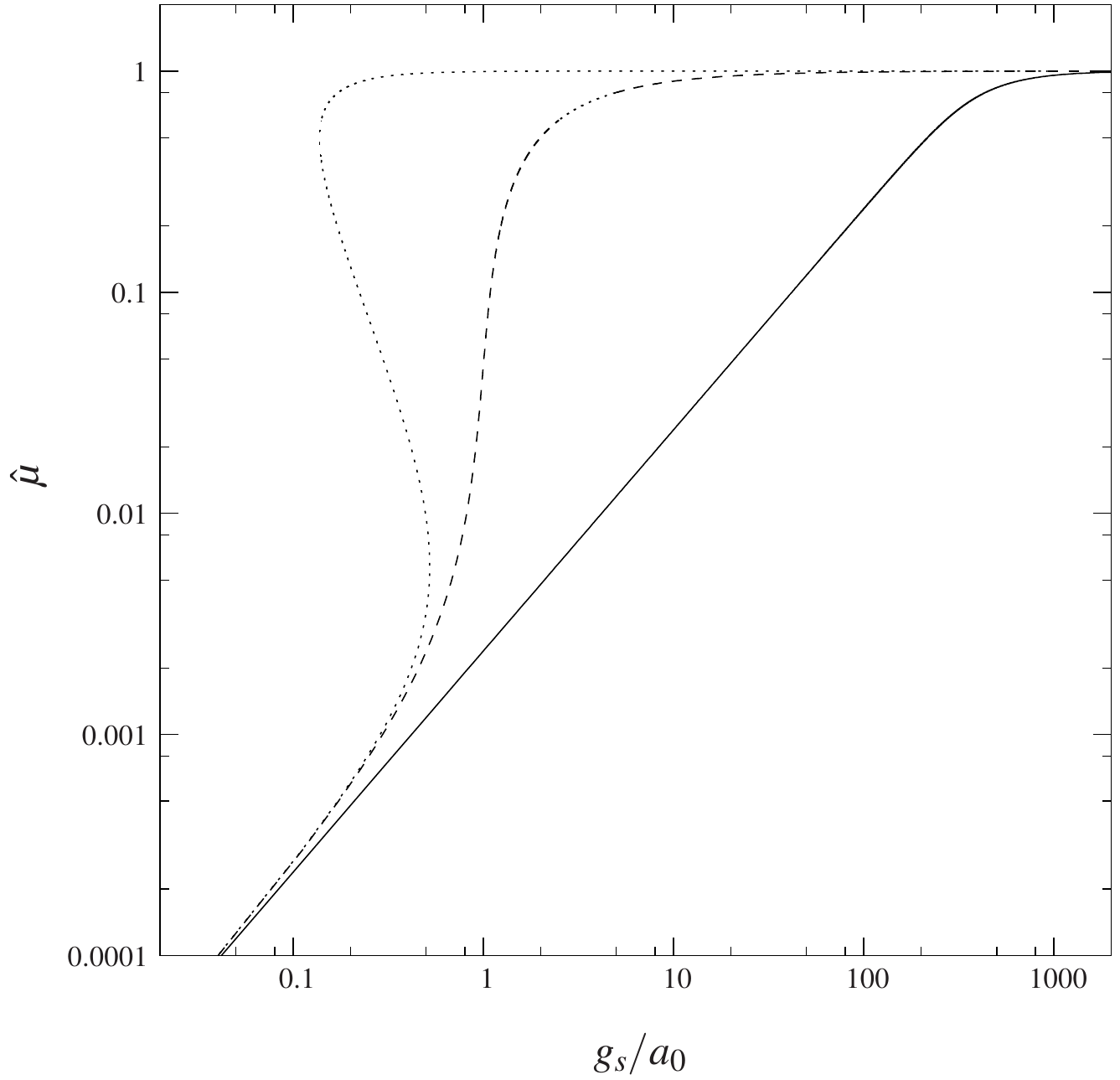}
\caption{Formally constructed TeVeS interpolating functions $\hat{\mu}$ for $k=0.03$ corresponding to the $\nu$-functions given by Eq.
\eqref{eq:m11} (solid line), Eq. \eqref{eq:m10} (dashed line) and Eq. \eqref{eq:m12} (dotted line).}
\label{fig:muteves}
\end{figure}

So far we have only considered the situation of low-acceleration bubbles within the QMOND framework which emerges in the nonrelativistic limit of particular
bimetric theories (see Sec. \ref{section2}). Another possibility of recovering MONDian dynamics in this limit is realized in TeVeS. In the following, we want
to address whether there exist viable interpolation functions for the TeVeS scalar field $\phi$ which give rise to similar results as found in Sec. \ref{section4}.

Applying the approximations for weak fields and quasi-static systems, the gradient of the total gravitational potential in TeVeS approximately takes the form
\cite{bekenstein2004,feix2008}
\beq
\bm{\nabla}\Phi = \bm{\nabla}\Phi_{N} + \bm{\nabla}\phi,
\label{eq:b1}
\eeq
where $\Phi_{N}$ is the Newtonian potential, obeying the usual Poisson equation, and the scalar field $\phi$ is a solution of
\beq
\bm{\nabla}\cdot\left (\hat{\mu}\bm{\nabla}\phi\right ) = kG\rho.
\label{eq:b2}
\eeq
Here $\rho$ denotes the density of real matter and $\hat{\mu}$ depends on the scalar field strength $g_{s}\equiv \lvert\bm{\nabla}\phi\rvert$
\footnote{Since we are working within the nonrelativistic approximation, we are only concerned with the quasistatic branch of the free
function}. In situations where the curl field vanishes, it follows from Eq. \eqref{eq:b2} that
\beq
\bm{\nabla}\phi = \frac{k}{4\pi\hat{\mu}}\bm{\nabla}\Phi_{N}
\label{eq:b3}
\eeq
which, if combined with Eq. \eqref{eq:b1}, leads to
\beq
\bm{\nabla}\Phi = \left (1+\frac{k}{4\pi\hat{\mu}}\right )\bm{\nabla}\Phi_{N}.
\label{eq:b4}
\eeq
Comparing the above with Eq. \eqref{eq:m1}, we identify the QMOND interpolating function $\nu$ as
\beq
\nu = 1 + \frac{k}{4\pi\hat{\mu}}.
\label{eq:b5}
\eeq
Together with our previous assumption that $\nu$ approaches unity for strong gravitational fields, the above leads to $\hat{\mu}\rightarrow\infty$ in
this limit. Such interpolation functions $\hat{\mu}$ have been found to be problematic and may therefore be regarded as inapplicable \cite{fmreview2011}.
Avoiding this issue by imposing $\hat{\mu}\rightarrow 1$ for $g_{s}/a_{0}\gg 1$, we obviously must have that $\nu\rightarrow 1+k/(4\pi )$ in the Newtonian
limit, consistent with Eq. \eqref{eq:b5} and resulting in an effective rescaling of the gravitational constant. Since we are here interested in the purely
non-Newtonian part of the gravitational potential, this can be easily achieved by formally adding the constant $k/(4\pi )$ to a given expression for
$\nu(y)$. As a consequence, the full solutions of Eq. \eqref{eq:m1} for the such constructed $\nu$-functions only differ by the term $k\Phi_{N}/(4\pi )$
from their counterparts, which leads to the same results as in Sec. \ref{section4} after subtracting the total Newtonian contribution. Another way of
saying this is the following: Adding a constant to $\nu(y)$ does not alter the effective density field if the region of interest is devoid of any real
physical matter sources. Choosing the same boundary conditions for the potential then yields identical results.
Keeping this technical detail in mind, we are now ready to construct the corresponding functions $\hat{\mu}$ with the help of Eqs. \eqref{eq:b3} and
\eqref{eq:b5}. Starting with the interpolation function Eq. \eqref{eq:m11}, a bit of algebra reveals that
\beq
\frac{\hat{\mu}}{\sqrt{1-\hat{\mu}^{4}}} = \frac{k}{4\pi a_{0}}g_{s},
\label{eq:b6}
\eeq
which is exactly the implicit definition of $\hat{\mu}$ used in Refs. \cite{bekenstein2006,bevis2010}. Similarly, one may find expressions for
the interpolating functions given by Eqs. \eqref{eq:m10} and \eqref{eq:m12}. Assuming $k=0.03$, Fig. \ref{fig:muteves} shows the resulting
TeVeS interpolation functions $\hat{\mu}$ for $0.1<g_{s}/a_{0}\lesssim 10^{4}$. While all of them exhibit the same asymptotic behavior, the
resulting $\hat{\mu}(g_{s}/a_{0})$ for Eq. \eqref{eq:m12} (dotted line) is multivalued, and therefore ill-defined \cite{fmreview2011,zhao2006}.
Since it is unclear how to assign $\hat{\mu}$-values to a given $g_{s}$ in this case, the boundary value problem in Eq. \eqref{eq:b2} turns
ambiguous, rendering the theory unphysical. The only remedy is to resort to a larger renormalization of the gravitational constant, i.e. a larger
$k$, but this would result in unacceptable cosmological predictions \cite{magueijo2011}. The interpolating functions constructed from Eqs. \eqref{eq:m10}
and \eqref{eq:m11} (dashed and solid lines) do not suffer from this problem and lead to healthy theories. The corresponding TeVeS stress signals
will be comparable to the QMOND results, but slightly lower due to a softening caused by the curl field \cite{magueijo2011}, which allows one to
adopt the conclusions of Sec. \ref{section4} for these cases.

\bibliography{references}

\end{document}